\documentclass[fleqn,usenatbib]{mnras}

\usepackage{newtxtext,newtxmath}

\usepackage[T1]{fontenc}
\usepackage{ae,aecompl}

\usepackage{graphicx}	
\usepackage{amsmath}	
\usepackage{algorithm2e}
\usepackage[version=4]{mhchem} 
\usepackage{float}
\usepackage{lscape}

\newcommand{\kms}{km\,s$^{-1}$} 

\title[Identification of molecular clouds in CHIMPS]{Identification of molecular clouds in emission maps:\\
a comparison between methods in the \ce{^{13}CO}/\ce{C^{18}O} ($J=3-2$) Heterodyne Inner Milky Way Plane Survey}

\author[R. Rani et al.]{
Raffaele Rani,$^{1}$\thanks{E-mail: rani@ntnu.edu.tw (RR)}
Toby J. T. Moore,$^{2}$
David J. Eden, $^{2,3}$
Andrew J. Rigby, $^{4}$
\newauthor
\  Ana Duarte-Cabral, $^{4}$
Yueh-Ning Lee  $^{1}$
\\
$^{1}$Center of Astronomy and Gravitation, Department of Earth Sciences, National Taiwan Normal University, 88, Sec. 4, Ting-Chou Rd., Wenshan District, \\ Taipei 116, Taiwan R.O.C.\\
$^{2}$Astrophysics Research Institute, Liverpool John Moores University, IC2, Liverpool Science Park, 146 Brownlow Hill, Liverpool, L3 5RF, UK\\
$^{3}$Armagh Observatory and Planetarium, College Hill, Armagh, BT61 9DB, UK\\
$^{4}$ School of Physics and Astronomy, Cardiff University, Queen's Building, The Parade, Cardiff, CF24 3AA, UK \\}

\date{Accepted XXX. Received YYY; in original form ZZZ}

\pubyear{2022}

\begin{document}
\label{firstpage}
\pagerange{\pageref{firstpage}--\pageref{lastpage}}
\maketitle

\begin{abstract}

The growing range of automated algorithms for the identification of molecular clouds and clumps in large observational datasets has prompted the need for the direct comparison of these procedures. However, these methods are complex and testing for biases is often problematic: only a few of them have been applied to the same data set or calibrated against a common standard. We compare the Fellwalker method, a widely used watershed algorithm, to the more recent Spectral Clustering for Interstellar Molecular Emission Segmentation 
(SCIMES). SCIMES overcomes sensitivity and resolution biases that plague many friends-of-friends algorithms by recasting cloud segmentation as a clustering problem. Considering the \ce{^{13}CO}/\ce{C^{18}O} ($J = 3 - 2$) Heterodyne Inner Milky Way Plane Survey (CHIMPS) and the CO High-Resolution Survey (COHRS), we investigate how these two different approaches influence the 
final cloud decomposition. Although the two methods produce largely similar statistical results over the CHIMPS dataset, FW appears prone to over-segmentation,
especially in crowded fields where gas envelopes around dense cores are identified
as adjacent, distinct objects.  FW catalogue also includes a number of fragmented
clouds that appear as different objects in a line-of-sight projection. In
addition, cross-correlating the physical properties of individual sources between catalogues is complicated by different definitions, numerical implementations, and design choices within each method, which make it very difficult to establish a one-to-one correspondence between the sources.

\end{abstract}



\section{Introduction}
The distribution and properties of gas within molecular clouds regulate, in part, the characteristics of newly formed stars, their numbers and masses, and the location of star-forming sites. The connection between the features of molecular
gas and both the initial mass function and formation rate of new stellar
populations have prompted a wide range of theoretical and observational studies
geared towards the characterisation of the structure of molecular clouds.
Multi-tracer surveys have revealed the hierarchical nature of these structures,
showing how high-density, small-scale features are always nested within more
rarefied, larger envelopes \citep{Blitz1986, Lada1992}.
This structural hierarchy is, however, a non-trivial one: at any scale, there
appear to be more high-density and compact `clumps' than larger and less dense
structures. The densest clumps in a cloud's hierarchy are compact cores, the
seeds of star formation. In these regions, over scales of about 0.1\,pc,
the turbulence in the cloud often becomes dominated by thermal
motions \citep{ Goodman1998, Tafalla2004, Lada2008}. The physical conditions
inside the cores determine the mechanisms involved in the conversion of
molecular gas into stars \citep{diFrancesco2007, WardThompson2007,
Bigiel2008,Schruba2011, Urquhart2018}. At the bottom of the density hierarchy, 
lie the low-density envelopes that surround the denser regions. 

The natural clumpiness that characterises the molecular
phase of the interstellar medium on different scales has led to the 
cataloguing of molecular emission by dividing the interstellar gas into
independent, discrete entities. Although this separation provides a useful theoretical distinction between giant molecular clouds and the diffuse
multi-phase interstellar medium, it is still unclear whether the density
hierarchy continues past this chemical boundary \citep{Blitz2007} extending
into the diffuse ISM \citep{BallesterosPaderes1999, Hartman2001}. In this
picture, the molecular phase of the ISM would not be enough to define the bottom
of the density hierarchy needed to treat a molecular cloud as an independent,
separate entity. 

Structural patterns in molecular emission have been investigated through a wide range of analysis methods.
Each technique focuses on the analysis of a different feature of the gas.  Fractal analysis \citep{Stutzki1998}, the study of power spectra \citep{Lazarian2000} and the structure function \citep{Heyer2004} have aimed to characterise turbulence in clouds \citep{Brunt2010, Brunt2014}, and clump identification algorithms \citep{Stutzki1990, FW, scimes} have been used to probe geometry, structure and substructure, e.g., the density hierarchy. In general, statistical approaches to the analysis of molecular-line data either aim to provide a statistical description of the emission over the entire dataset or a division of the emission into physically relevant features. The latter approach is then followed by the analysis of the characteristics of the resulting population of sources.
Statistical approaches include fractal analysis \citep{Elmgreen1996, Stutzki1998, Elmegreen2002, Sanchez2005, Lee2016}, $\Delta$-variance \citep{Stutzki1998, Bertram2015}, correlation functions \citep{Houlahan1990, Rosolowsky1999, Lazarian2000, Padoan2003} and analysis of the two-dimensional power spectrum \citep{Schlegel1998, Pingel2018,Combes2012, Feddersen2019} and principal components \citep{Heyer2004}. 
These techniques provide the overall statistical properties of the sample and are thus best suited for the comparison of measurements between different datasets. On the other hand, clump identification (image segmentation) is preferred for the study of physically important substructures embedded in the emission. In position-position velocity (PPV) data sets, giant molecular clouds (GMCs) and their substructure are identified as discrete features (sets of connected voxels) with emission (brightness temperature or column densities) above a specified threshold \citep{Scoville1987, Solomon1987}. 

Molecular-cloud recognition in PPV data sets is performed with a variety of automated algorithms. These methods are commonly designed to operate on large data sets and different levels of blending between structures. Two different strategies for the identification of molecular emission are frequently employed in the construction of GMC identification software packages:
the iterative fitting and subtraction of a given model to the molecular emission \citep{Stutzki1990, Kramer1998} and the friends-of-friends paradigm that connects pixels based on their and their neighbours' emission values \citep{Williams1994, Rosolwsky2006}. The latter approach is often applied as a watershed formulation in which single objects are identified as partitions of the data corresponding to sets of paths of steepest descent around signal peaks. This strategy thus recasts GMC recognition as an image segmentation problem \citep{Pal1993}. Contouring in three-dimensional images, however, remains a complex task. Complications arise from the difficult deblending of internal structures in crowded regions as the boundaries that separate star-forming clouds from the surrounding multi-phase ISM are often unclear \cite[see][]{BallesterosPaderes1999, Hartman2001, Blitz2007}.  The efficacy of GMC recognition is thus affected by survey-specific biases arising from spatial and spectral resolution and the sensitivity in molecular-line observations of GMCs  \citep{Rosolwsky2006, Pineda2009, Wong2011}. Cloud recognition usually worsens in regions characterised by complex molecular environments and crowded velocity fields  (such as the Inner Milky Way), where resolution plays a crucial role in the identification
of structure \citep{Hughes2013}. At low resolution, segmentation algorithms suffer from the blending of emission from unrelated clouds \citep{Colombo2014}, while high resolutions cause cloud substructures to be identified as individual clouds. In particular, friends-of-friends methods are especially sensitive to 
resolution. In clumpy environments, the objects naturally selected by this type of algorithm have the scale of a few resolution elements \citep{Rosolwsky2006}.


Recently, alternative segmentation methods based on the physical properties of molecular gas have been proposed, most noticeably gravitational acceleration mapping methods \citep{Li2015} and dendrograms \citep{Rosolowsky2008}. Dendrograms are particularly well-suited to encode the essential features of the hierarchical structure of the isosurfaces for molecular line data cubes. They represent the changing topology of the isosurfaces as a function of contour level. This growing range of automated cloud-identifying paradigms and their implementations has prompted the need for a direct comparison of the methods. However, the algorithms are often complex and testing for biases is not straightforward as only a few of them have been applied to the same data set or calibrated against a common standard \citep{Lada2020}. 

Although the performance of several popular clump-finding algorithms has recently been compared on artificial emission maps \citep{Li2020}, cross-correlating the physical properties of individual sources between several catalogues is a non-trivial task. From this viewpoint, it is thus useful to apply different methodologies to identify and extract GMCs from the same survey. 
In this study, the Spectral Clustering for Interstellar Molecular Emission Segmentation (SCIMES) algorithm is applied to
identify GMCs in the \ce{^{13}CO}  data-set of the  \ce{^{13}CO}/\ce{C^{18}O}($J=3 - 2$) Heterodyne Inner Milky
Way Plane Survey (CHIMPS). To directly compare this segmentation to the results obtained by
\citet{Rigby2019} with the FellWalker (FW) algorithm, the dendrogram defining parameters
are chosen to match the FW input configuration. SCIMES makes use of dendrograms to encode the hierarchical
structure of molecular clouds and then employs spectral clustering to produce dendrogram cuts corresponding to the
individual clouds \citep{scimes}, whereas FW is a variation of the watershed paradigm, based on the paths of steepest ascent \citep{FW}. To extend the comparison to the properties of a different tracer, to show the effect of isotopologue choice, a SCIMES segmentation of the $\ce{^{12}CO} (3-2)$ emission from the CO High-Resolution Survey (COHRS; \citealt{Dempsey2013}) is considered on the regions covered by CHIMPS. 

We present an empirical comparison between the FW and SCIMES algorithms on a large sample of clouds within the 
 \ce{^{13}CO}/\ce{C^{18}O} ($J = 3 - 2$) Heterodyne Inner Milky Way Plane Survey (CHIMPS). To do so, 
we construct a novel catalogue of CHIMPS sources obtained through the application of SCIMES. The catalogue includes a number of measured and calculated cloud properties chosen to match those defined in \cite{Rigby2019}.

In Section \ref{data}, we briefly describe the CHIMPS data used in our analysis. A description of a SCIMES
source extraction that matches the FW parameterisation is provided in Section \ref{source_extraction} and the subsequent distance
assignments in Section \ref{distance_assignments}. Section \ref{results} presents a statistical comparison of the salient
physical properties of the sources in the SCIMES and FW catalogues, while Section \ref{Conclusions} summarises and discusses the results found in this study.

\section{Data}
\label{data} 

The \ce{^{13}CO}/\ce{C^{18}O} ($J=3-2$) Heterodyne Inner Milky Way Plane 
Survey (CHIMPS)  is a spectral survey of the  $J = 3 - 2$ 
rotational transitions of \ce{^{13}CO} at 330.587\,GHz and \ce{C^{18}O} at 
329.331\,GHz. The survey covers $\sim$19 square degrees of the Galactic plane, spanning longitudes $l$ between  $27\fdg 5$ and $46\fdg 4$ and latitudes $|\, b \,| < 0\fdg5$, with angular resolution of 15\,arcsec. The 
observations were made over a period of 8 semesters (beginning in the spring of 2010) at the $15$-m James Clerk Maxwell Telescope (JCMT) in Hawaii. Both 
isotopologues were observed concurrently \citep{Buckle2009} using the 
Heterodyne Array Receiver Programme (HARP) together with the Auto-Correlation Spectral Imaging System (ACSIS). The data obtained are organized in position-position-velocity (PPV) cubes with 
velocities binned in 0.5 \kms\ channels and a bandwidth of 
200\,\kms. 

The Galactic velocity gradient associated 
with the spiral arms (in the kinematic local standard of rest, LSRK) is matched by shifting the velocity range with increasing Galactic longitude, as observed in previous molecular Galactic plane studies \citep[e.g.][]{Dame2001}. Varying the range from $-50 < v_{\rm LSR} < 150$\,\kms\ at $28^\circ$ to $-75 < v_{\rm LSR} < 125$\,\kms\ at $46^\circ$, we recover the 
expected velocities of objects observed in the Scutum-Centaurus tangent and the 
Sagittarius, Perseus and Norma arms. 

The \ce{^{13}CO} survey has mean rms 
sensitivities of $\sigma(T_{\rm A}^{*})\approx 0.6$\,K per velocity channel, while for 
\ce{C^{18}O}, $\sigma(T_A^{*})\approx 0.7$\,K, where $T_A^{*}$ is the antenna 
temperature corrected for atmospheric attenuation, ohmic losses inside the instrument, spillover, and rearward scattering \citep{Rigby2016}. 
These values, however, fluctuate across the survey region depending on both weather conditions and the varying numbers of working receptors on HARP. In \ce{^{13}CO} (3–2), the rms of
individual cubes ranges between  $\sigma(T_A^{*}) = 0.37$ K and 1.51 K per channel, and between  $\sigma(T_A^{*}) = 0.43$ K and 1.77 K per channel in \ce{C^{18}O} (3–2) \citep{Rigby2016}. 

Column density maps are necessary for the estimation of the cloud masses (see Section \ref{mass}). The total column densities throughout the CHIMPS survey were calculated from the excitation temperature and the optical depth of the CO emission. This calculation is outlined in 
\citet{Rigby2019}. Their method is a variation 
of the standard calculation of the excitation temperature and optical 
depth \citep{Wilson2013} and uses the \ce{^{13}CO}($J=3-2$) emission at each 
position $(l,b,v)$ in the datacube on a voxel-by-voxel 
basis under the assumption of local thermodynamic equilibrium. 
The major advantage of this strategy over the analysis of 
velocity-integrated properties is that any property derived from the 
excitation temperature and optical depth is independent of source 
extraction and image segmentation algorithms. However, individual voxel information does not account for the attenuation of the emission due to 
self-absorption along the line of sight. \cite{Rigby2019} performed a first-order adjustment of the method with respect to the \ce{^{12}CO}($3-2$) from which the excitation temperature of \ce{^{13}CO}($3-2$) is derived and did not find evidence for significant self-absorption in \ce{^{13}CO}($3-2$) across the entire CHIMPS survey. The total column density at each position is determined from the column density within a specific energy level by multiplication with an appropriate partition function representing the sum over all states 
\citep{Rigby2019}. 

COHRS mapped the \ce{^{12}CO} ($3-2$) emission in the Inner Milky Way plane, covering latitudes $10 \fdg 25 < l < 17\fdg 5$ with longitudes $|\,b\,| \leq 0\fdg 25$ and $17\fdg 5 < l < 50\fdg 25$  with $|\,b\,| \leq 0\fdg 25$. 
This particular region was selected to match a set of
important surveys, among which are CHIMPS, the Galactic
Ring Survey \cite[GRS;][]{GRS}, the FOREST Unbiased
Galactic plane Imaging survey with the Nobeyama 45-m
telescope  \cite[FUGIN; ][]{FUGINI}, the Galactic Legacy
Infrared Mid
 Plane Survey Extraordinaire  
\cite[GLIMPSE;][]{GLIMPSE}, the Bolocam Galactic Plane Survey \cite[BGPS; ][]{bolocam}, and the {\em Herschel} Infrared Galactic Plane Survey \cite[Hi-GAL; ][]{higal}. The observations were  also performed at JCMT with HARP at $345.786$\, GHz and ACSIS set at a 1-GHz bandwidth yielding a frequency resolution of 
$0.488$\,MHz ($0.42$\,km\,s$^{-1}$). The survey covers a velocity range between  $-30$ and $155$\,\kms, with a spectral resolution of $1$\,\kms and angular resolution of $16.6$\, arcsec (FWHM). 
The COHRS data (first release) are publicly available\footnote{\url{http://dx.doi.org/10.11570/13.0002}}. We consider a sub-sample of the full set of COHRS sources by only considering those within the regions covered by CHIMPS.

\begin{figure*}
	\includegraphics[width=\textwidth]{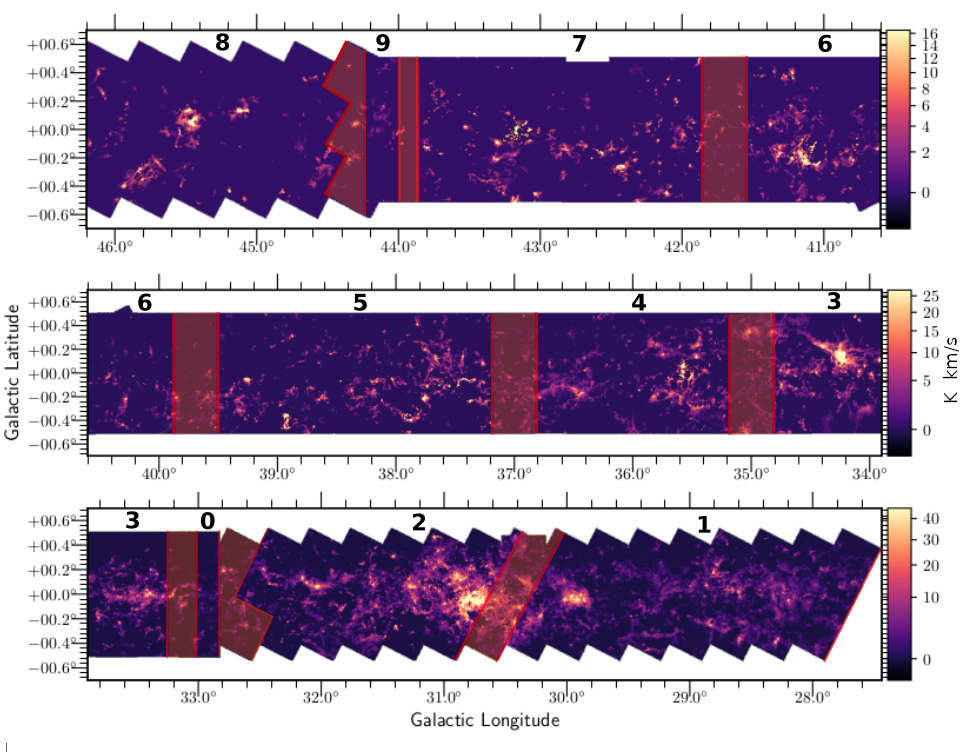}	
	\caption{Integrated intensity map ($\int T_A^* dv$) of CHIMPS (full survey). The colour bar shows the scaling in units of K\,\kms. The 10 regions into which the survey is divided are delimited by red lines. Orange shading denotes the overlapping areas between adjacent regions. Region numbers are printed above the map.}
	\label{rcuts} 
\end{figure*}

In this analysis of the difference between the FW and SCIMES extraction algorithm, we consider the 
($J = 3 - 2$) emission from the reduced data in the 10 regions constituting the CHIMPS survey (Fig.\,\ref{rcuts}).  To directly compare the new SCIMES segmentation to the results obtained with the FW algorithm by \citet{Rigby2019},  the dendrogram-defining parameters are chosen to match the FW input configuration as closely as possible, as described in the next Section.

\section{Source extraction}
\label{source_extraction} 

We use the 
SCIMES algorithm, first introduced by \citet{scimes, Colombo2018}, to decompose the \ce{^{13}CO} emission into 
individual molecular clouds (sources). SCIMES is a publicly available Python package that uses spectral clustering to identify single objects within a dendrogram that represents the hierarchical structure of the emission \citep{Rosolowsky2008}. 
The emission dendrogram is produced using the Python package for astronomical dendrograms \cite[Astrodendro, ][]{astropy:2013, astropy:2018}. In the framework of SCIMES, the leaves of the dendrogram are identified with the local maxima in the emission and the branches represent isosurfaces (contours in the PPV data) at different emission levels (they are structures containing other branches and leaves). 

SCIMES uses similarity criteria to analyze a dendrogram by translating it into a weighted complete graph. In the associated graph, the vertices correspond to the leaves in the dendrogram and weights on the edges encode the affinity relationship between the leaves (larger values of the affinity represent the higher similarity between two vertices of the graph).
The SCIMES algorithm then uses spectral clustering on the affinity matrix representing the graph to partition the graph into separate components. These clusters define a segmentation of the emission into individual clouds. 
This process partitions the graph into $k$ regions, which coincide with the molecular emission features encoded by the dendrogram and consequently to the connected regions of the emission in PPV space.  These `molecular gas clusters' are labelled as clouds, clumps, or cores depending on the scale of the emission. As the SCIMES decomposition considers the natural transitions in the emission structure to segment PPV data and is robust across scales, it has the major advantage of being applicable to a variety of spatial dynamic ranges \citep{scimes}. 

Because of the variable weather conditions and the varying number of 
active receptors during the 4 years of observations, the original CHIMPS 
datacubes do not present a completely uniform sensitivity across the entire survey \citep{Rigby2016}. To avoid loss of good signal-to-noise sources in regions of low background and to prevent high-noise regions from being incorrectly identified as clouds, we perform the source extraction on the signal-to-noise ratio (SNR) cubes instead of brightness-temperature data. This approach was applied to continuum data in the JCMT Plane Survey \citep{Moore2015, Eden2017}, who noted that this method produced the best extraction results. We define the SCIMES parameters as multiples of the background $\sigma_\textrm{rms}$. For signal-to-noise cubes, $\sigma_\textrm{rms}=1$ by definition. 

The reduced data are organised into 178 datacubes which are, in turn, mosaiced into 10 larger regions (Fig. \ref{rcuts}) since the entire CHIMPS area is too large to be analysed as a single datacube. For each region, we set the SCIMES parameters to generate a dendrogram of 
the emission in which each branch is defined by an intensity change ($\mathtt{min\_delta}$) of $5\sigma_{\textrm{rms}}$ and contains at least three resolution elements worth of pixels ($\mathtt{min\_npix}=16$). Any emission below $3\sigma_\textrm{rms}$ ($\mathtt{min\_val}=3\sigma_\textrm{rms}$) is not considered. These specific values were chosen to match the corresponding {\sc FellWalker} configuration parameters $\mathtt{FellWalker.MinHeight}$, $\mathtt{FellWalker.Noise}$, $\mathtt{FellWalker.MinPix}$ \citep{FW} used by \citet{Rigby2016} for their CHIMPS extraction.


The emission dendrogram is produced using the Python package for astronomical dendrograms \cite[{\sc Astrodendro}, ][]{astropy:2013, astropy:2018}. However, the {\sc Astrodendro} implementation that SCIMES uses to construct the emission dendrogram does not make a distinction between the spatial and spectral axes. Thus, some clouds that are unresolved in one 
dimension may still be included in the dendrogram. These sources are 
eliminated in a post-processing step. Since the distance distance assignments to the dendrogram structures cannot be made before the full segmentation (see Section \ref{distance_assignments}), we cannot generate the volume and luminosity affinity matrices required for spectral clustering from spatial volumes and intrinsic luminosities. Instead, we use PPV volumes and integrated intensity values. 

In addition, we retain single leaves that do not form clusters (since clusters are constituted by at least two objects) and the (sparse) clusters constituted by the intra-clustered leaves \citep{Colombo2015}, which are usually discarded as noise \citep{Ester1996}. This way the SCIMES algorithm behaves as a 'clump finder' \footnote{\url{https://scimes.readthedocs.io/en/latest/tutorial.html}}. Although this choice allows for the segmentation to include sources that cannot strictly be defined as 'molecular gas clusters' \citep{Colombo2015}, these clouds are expected to match the clumps found in the BGPS \citep{bolocam}.

\subsection{Post-processing filter}
\label{sec:post_processing}

To clean the catalogue of spurious sources and noise artefacts that are left after extraction, we apply an additional filter. This filter leaves those clouds that extend for more than 3 voxels in any direction (spatial or spectral). While the fhttps://g.co/verifyaccountormer requirement ensures that we are considering thin filaments, the latter ensures that each cloud is fully resolved in each direction (the width of the beam being 2 pixels). In addition, we remove a number of smaller clouds in contact with the edges of 
the regions and those 6voxels which lack a column density assignment. Although the segmented structures are mostly coherent, different velocity components may sometimes be blended in the same object in clouds associated with the border of the field of observation, since these sources do not present
closed contours. Finally, to construct the final catalogue and its corresponding assignment mask, we apply a selection criterion to handle the clouds in overlapping areas between adjacent regions. This procedure is described below.

\subsection{Overlapping areas}
\label{overlapping_regions}

\begin{figure}
 \includegraphics[width=\columnwidth]{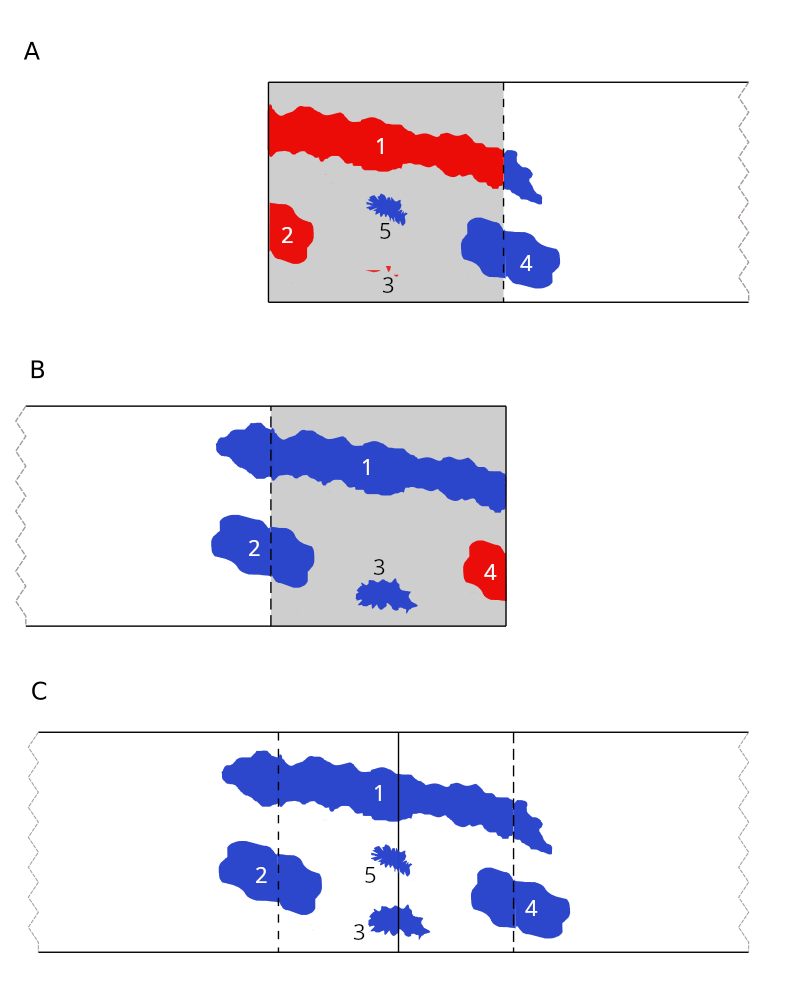}
 \caption{Prescription for cloud removal in the overlapping area (shaded area in the panels) of adjacent regions (panels A and B). In each region, we remove the clouds within the overlapping areas that cross longitudinal edges. The clouds and parts portions of clouds that are removed in each region (clouds 1, 2, and 3  in panel A, and 4 in panel B) are drawn in red. These sources are recovered from the adjacent region. Clouds that span the entire overlapping area (cloud 1) are split at the longitudinal edge that marks the end of the region (panel A). The portion of the cloud contained in the shaded area is then assigned to the region that contains most of the cloud (panel B) and removed from the other (panel A). The portion of the cloud left in panel A (blue tip) is then added to the final catalogue (panel C). Whenever two (or more) clouds overlap (cloud 3), we discard the smaller object between the two regions. We retain all objects  that  do  not  overlap  between  the  regions (cloud 5).}
 \label{fig:edges}
\end{figure}

Each of the 10 regions into which CHIMPS is divided contains a variance array component determined for each spectrum from the system noise temperature. In order to perform source extraction as
consistently as possible, a small overlap is left between adjacent regions. To avoid double-counting clouds and to account for the 
discrepancies in the extraction maps near longitudinal 
edges due to the separate dendrograms representing the 
gas structure in each region, we use the following 
prescription to treat sources extracted in the 
overlapping areas. In each region, we remove clouds 
within the overlapping area that cross the longitudinal 
edges of the region (clouds 2 in panel A and 4 in panel B of Fig.~\ref{fig:edges}). Such 
clouds do not have closed isocontours in the region in 
question \citep{scimes}. We recover these objects from 
the SCIMES extraction in the adjacent regions, which 
contain the clouds to their full extent. Some regions 
present clouds that span the entire overlapping field. 
In order not to discard a significant amount of gas 
mass, we split these clouds at the edge of one region, 
assigning the portion in the overlapping area to the 
region that contains most of the cloud (cloud 1 in panels A and B in Fig.~\ref{fig:edges} becomes assigned to the region depicted in panel B).
The remaining portion of the cloud, left in the adjacent region, is 
then added to this catalogue entry, considering its distance as the same 
as the distance of the larger part. Since this situation occurs for one 
source only in the entire catalogue (between regions 3 and 4), the physical 
properties of this source were calculated manually 
taking into account the properties of the voxels in each region and making 
the required adjustments. 
Finally, we include  all  objects  
that  do  not  overlap  between  the  regions (cloud 5 
in Fig.~\ref{fig:edges}), and whenever two (or 
more) clouds overlap, we simply discard the smaller 
object between the two regions (cloud 3 in panel A 
in Fig.~\ref{fig:edges}). Through this procedure,
we construct a catalogue of 2944 molecular clouds.

Finally, to produce a fair comparison of the physical properties of 
clouds, we match the FW subcatalogue by only considering SCIMES sources 
that contain at least a voxel with $S/N \geq 10$ \citep{Rigby2019}. Thus, the final SCIMES catalogue used in the analysis that follows amounts to 1586 sources. None of the sources
left after this selection is a single isolated leaf.

\section{Distance assignments}
\label{distance_assignments} 

A distance assignment to the extracted SCIMES sources was constructed by combining different catalogues and using a Bayesian distance estimator \citep{Reid2016}. 
We first consider the latest version of the ATLASGAL source catalogue \citep{Urquhart2018}.
Distances were assigned as follows. Each SCIMES cloud is matched to a set of one or more ATLASGAL sources. The matching process is performed through an area ($l$,$b$) search that allows the closest sources (Euclidean metric) that lie within a neighbourhood of radius $r$ arcsecs centred at the centroid of the SCIMES object to be selected. The radius $r$ is taken by adding $38$\,arcsec ($\approx 5$\,pixels) to the radius of the SCIMES object (following \citealt{Rigby2019}). Next, if this search returns multiple clouds, the distance that most sources have in common is chosen. If the distances in the set vary significantly we check if any of them belongs to an ATLASGAL cluster, and assign the cluster's distance to the SCIMES cloud. SCIMES clouds that contain one single ATLASGAL source for which the distance is not available, or in the case of clusters, ATLASGAL does not provide a cluster distance, are left unassigned.  

We then consider a sub-catalogue of the FW assignments \citep{Rigby2019}. This subset of the FW sources comprises only robust sources. These are sources that are not false positives or single coherent sources at low S/N which are hard to discern by eye. The reduced catalogue is also free of sources consisting of diffuse gas at low S/N that may contain multiple intensity peaks, or irregular profiles (resulting from the segmentation of clouds across tile boundaries). This robust sub-catalogue amounts to 3664 entries. We will refer to this catalogue as the FW catalogue. 
The Bayesian distance calculator was used to estimate the possible near and far kinematic distance - and associated uncertainties - for each of the clumps \citep{Rigby2019}. No assumption about the sources being associated with spiral arms was made, and the standard Galactic rotation model \citep{Reid2014}, with a distance to the Galactic centre of $R_0= 8.34 \pm 0.16$\,kpc was adopted for the calculations.


SCIMES clouds without ATLASGAL counterparts are compared to the FW catalogue. If a SCIMES cloud contains a single FW object (emission peak) or more FW objects with the same distance, then that distance is assigned to the cloud. If a SCIMES cloud contains multiple FW sources with different distances, the distance that corresponds to the mode of the distribution of FW distances is assigned. If this distribution has no modes, the first FW source in the list is chosen.



Since the SCIMES and FW do present discrepancies in the emission structures they identify (see the small clumps at latitude smaller $-0.2^\circ$ in Fig. \ref{reg7} in Appendix \ref{exa}), not all the SCIMES clouds contain one or multiple FW. For the remaining unassigned clouds, associations between the unassigned SCIMES sources are made using a final volumetric search. This time an ellipsoidal volume of semi-axes $0\fdg3 \times 0\fdg3 \times 10$\,km\,s$^{-1}$, centred at the centroid of each remaining cloud, is employed to identify the closest SCIMES centroid with an existing distance assignment. The size of this volume is in agreement with the appropriate tolerance for friend-of-friends grouping \citep{Wienen2015} and corresponds to the median angular size and maximum linewidth of molecular clouds \citep{Romanduval2009}.

\begin{figure}
	\centering 
	\includegraphics[width=\columnwidth]{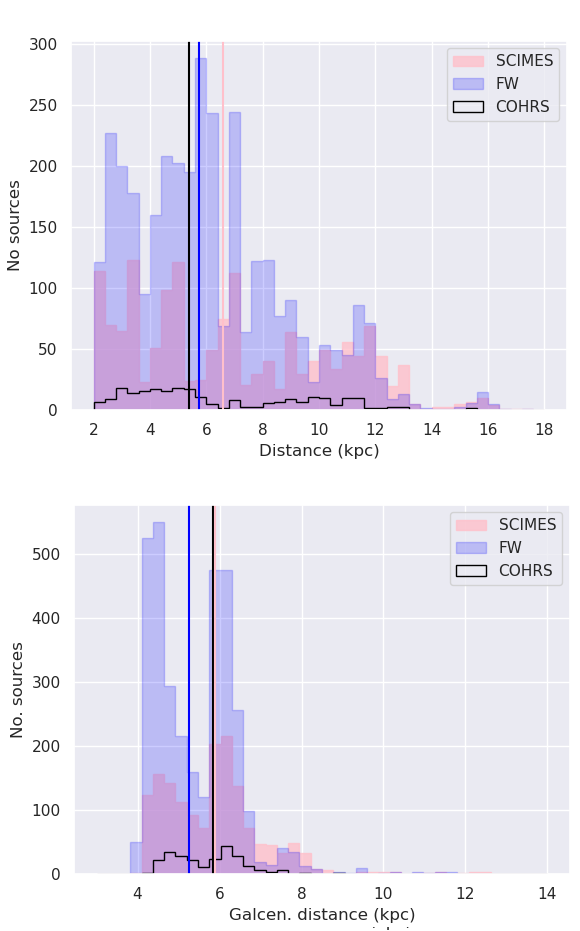} 
	\caption{Distributions of heliocentric and galactocentric distances for the CHIMPS \ce{^{13}CO} (3 - 2) sources extracted through the FW and SCIMES segmentations. The black histogram is the distribution of sources in a subset of the COHRS catalogue. The vertical lines denote the median values of the distributions. The median values of the distributions of heliocentric distance are $5.9$, $5.3$, and $5.8$ kpc for the SCIMES, FW, and COHRS source respectively. In the case of galactorcentric distances, the median values are $6.6$, $5.7$, and $5.4$ in SCIMES, FW, and COHRS respectively.}
	\label{dist} 
\end{figure}

\begin{figure}
    \centering
	\includegraphics[width=\columnwidth]{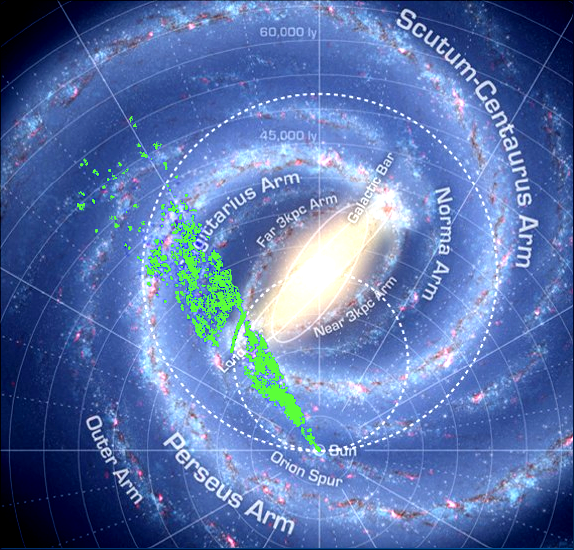}	
	\caption{Top-down view of the locations of the \ce{^{13}CO} (3 - 2) extracted through the
	SCIMES algorithm from CHIMPS. The background image is published by \citep{Churchwell2009}. The Solar circle and the locus of the tangent points have been marked as dashed and dotted lines respectively.}
	\label{galmap} 
\end{figure}

Finally, Reid's Bayesian calculator is employed to estimate the distances of the remaining SCIMES sources with undetermined distances with a near-far probability of $0.5$.

%

To avoid contamination of the results by local sources and to exclude 
a large number of low-luminosity clumps/clouds below the completeness limit, only sources with heliocentric distance $> 2$\,kpc are included \citep{Urquhart2018}.

Galactocentric distances are calculated independently through
\cite{Brand1993}'s rotation curves. The angular velocity is derived from the line-of-sight velocity, $v_\mathrm{LSR}$, and the Galactic coordinates $l$ and $b$ via the relation

\begin{equation}\label{rotation}
    \omega = \omega_0 + \frac{v_\mathrm{LSR}}{R_0 \sin(l)\cos(b)},
\end{equation}

\noindent
where $\omega_0 = 220$ km\,s$^{-1}$\,kpc$^{-1}$  is the Sun's angular velocity at its Galactocentric distance $R_0 = 8.5$\,kpc. The Galactocentric distance of a source is then obtained by solving

\begin{equation}\label{brand_galcen}
    \frac{\omega}{\omega_0} = a_1\bigg(\frac{R}{R_0}\bigg)^{a_2-1} + a_3\frac{R_0}{R}
\end{equation}

\noindent
numerically, with the constants $a_1= 1.0077$,  $a_2 =0.0394$, and $a_3=0.0071$ \citep{Brand1993}.

Fig.\,\ref{dist} shows the distribution of distances to CHIMPS \ce{^{13}CO} sources extracted with both FW and SCIMES. For comparison, the distance distribution of the subsample of COHRS sources is included.

The absence of a one-to-one correspondence between FW and SCIMES clouds makes it impossible to establish a unique matching criterion between the FW and SCIMES distance assignments of each cloud. In the assignment method described above, a distance is assigned to a SCIMES cloud based on the FW sources it contains. The difference in the numbers of clouds at large distances ($\sim 12$\,kpc) and at $\sim 5$\,kpc in the FW and SCIMES catalogues are a consequence of the differences in the segmentations and the assignment scheme of Section\,\ref{distance_assignments}. The larger number of clouds seen in the SCIMES catalogue at 12\,kpc arises from those assignments that do not involve FW distances. To check the robustness of the distance assignments great than 12 kpc, we consider the Larson relations and the galactic latitude of this set of sources (133).  
The Larson relations confirm the scaling obtained for the full sample 
discussed in Section \ref{scaling_relations}, while $50\%$ are off the Galactic plane with latitudes either smaller than $-0.15 ^{\circ}$ or greater than $+0.15^{\circ}$.

However, as we discuss below, when the statistical properties of a large ensemble of sources are considered, the impact of a particular choice of distance assignment on the derived parameters and properties for individual clouds becomes less prominent.

The top-down view of the locations of the CHIMPS sources 
extracted by SCIMES on the Galactic plane is shown in
Fig.\,\ref{galmap}. No sources closer than 3.5 kpc from the
Galactic Centre are found since the CHIMPS data do not
probe longitudes closer to the centre. The sources in our
sample reside within the four main spiral arms, the Scutum-Centaurus, Sagittarius-Carina, Perseus and Outer arms and
the smaller Aquila Rift and Aquila Spur features. The
spiral-arm structure is mirrored by the distribution of the
sources' Galactocentric distances.  The lower panel of
Fig.\,\ref{dist} displays large peaks at $\sim 4.5$ and $\sim 6.5$ kpc. These are the locations of the Scutum and Sagittarius arms seen from the Galactic Centre. The smaller peak at  $\sim 7.5$ kpc corresponds to the sources collected in the Perseus arm. As a section of the Scutum arm traverses the locus of tangential circular velocities, the sources in this area become clustered along this locus leaving gaps either on both sides (Fig.\,\ref{galmap}). This artefact stems from sources that have velocities greater than the terminal velocity due to non-circular streaming motions, which get binned at
exactly the tangent distance, resulting in the apparent ‘gap’ and arc of sources lying on the tangent circle.


\begin{figure*}
	\centering 
	\includegraphics[width=1.03\textwidth]{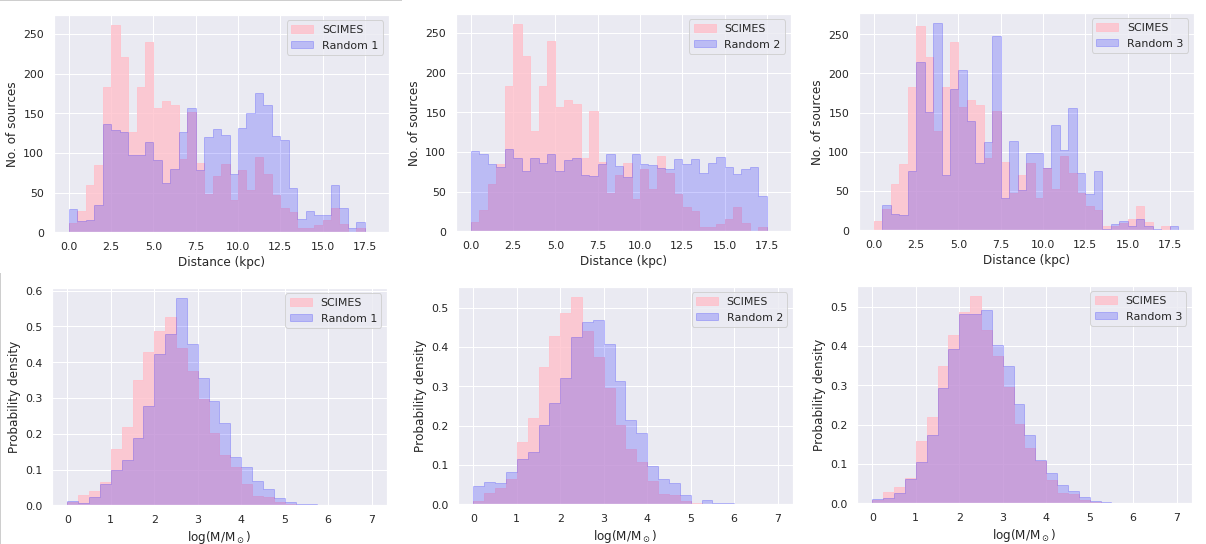} 
	\caption{Top row: distribution of the three sets of random distances compared to the assigned distances to SCIMES clouds in CHIMPS (SCIMES). From left to right: the first set (Random 1) corresponds to distances drawn from the set of unique distances that were assigned to SCIMES sources. The second set (Random 2) is drawn from the set of (equispaced) distances between the minimum and maximum value of the SCIMES distance. Finally, the set Random 3 is drawn from the distribution of distances generated from the original SCIMES. Bottom row: distribution of masses estimated from the random distance sets compared to the masses corresponding to the original SCIMES distance assignments.}\label{random1}
\end{figure*}

\subsection{A note on distances} \label{note} 

To quantify the impact of the choice of distance assignment on the physical properties of the clouds in the catalogue, we consider three 
random distance assignments and check their corresponding distributions
of masses. For the full SCIMES catalogue, the random distance assignments consist of applying a distance to each SCIMES cloud by drawing the value from 

\begin{enumerate}
    \item the set of all distances assigned to the SCIMES sources (each distance has the same probability of being assigned),
    \item a set of (equispaced) distances between the minimum and maximum value of the SCIMES distance assignments, 
    \item a probability distribution (weights) generated from the original distribution of distances
\end{enumerate}

The distance distributions derived from these assignments are compared to that of the original assignment in Fig.\,\ref{random1}. This figure also depicts the distributions of masses associated with the three random distance assignment methods described above. The masses corresponding to each random distance assignment were estimated as in subsection \ref{mass}. Although the distance distributions are largely dependent on the chosen assignment method, their differences are strongly mitigated when the corresponding masses are considered.

Performing the Kolmogorov-Smirnov test to check whether the original mass assignment and the random assignment are samples from the same distribution returns $k = 0.0797$ with p-value$=8.9931\times 10 ^{-8}$ 
and $k = 0.0790$ with p-value $=1.2360 \times 10 ^{-7}$ for distance drawn randomly from the original set of distance assignments and from 
a set of (equispaced) distances between the minimum and maximum value of the SCIMES distance assignments (see above). 
Finally. when we consider distances drawn from a probability distribution (weights) generated from the original distribution of assigned distances describe in Section \ref{distance_assignments}, the test returns $k = 0.0267$ with p-value $=0.2994$.

These results thus demonstrate that mass distribution obtained from randomly assigned distances is independent of the distribution of mass obtained with the distance assigned through the algorithm in Section \ref{distance_assignments} unless the values are randomly chosen from the original distribution of distances. Similar results can be obtained for other physical properties that depend directly on distance, e.g. cloud radii, area, and surface densities. 
Even though the purely random distributions show deviations in the statistics, we do not actually expect the actual distribution of distances to the observed clouds in this sample to differ much from the assigned one (the first two cases above are extremes). Thus this test shows us that inaccurate distances to clouds are not crucial when the overall population still follows the expected distance distribution. The size of a sample containing a wide range of cloud sizes and geometries thus mitigates the inaccuracies and differences arising from imprecise distance assignments.

\section{Comparison between FW and SCIMES segmentations}\label{results}

\begin{figure}
	\centering 
	\includegraphics[width=1.1\columnwidth]{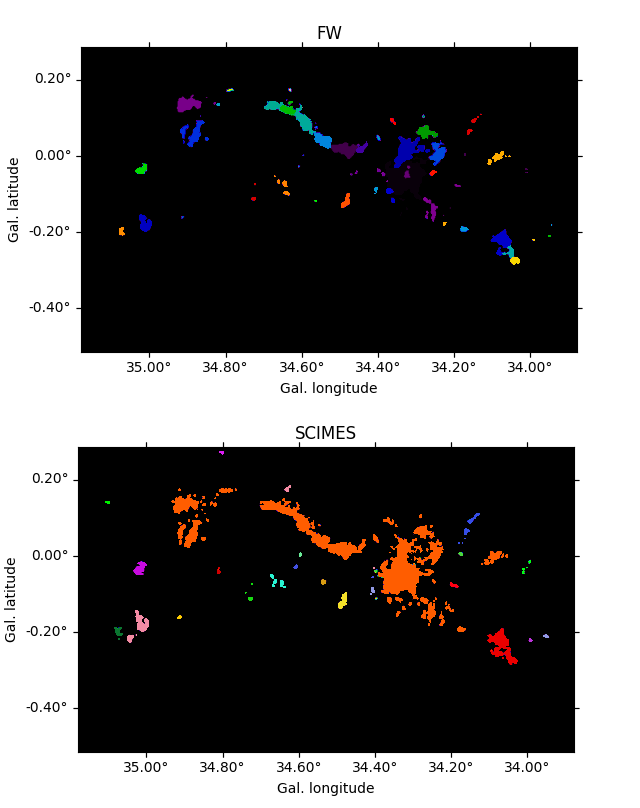} 
	\caption{Corresponding FW (top) and SCIMES (bottom) clusters in the 59.72-km\,s$^{-1}$ velocity plane at 27.4-arcsec resolution (see text). In both panels, different colours represent different clouds. 
	}
	\label{efs} 
\end{figure}
\begin{figure}
	\includegraphics[width=1.15\columnwidth]{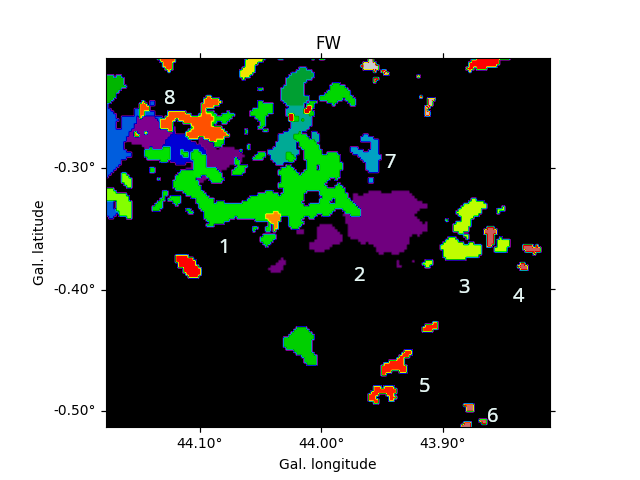} 
	\caption{Example of disconnected clouds in the FW
 segmentation. The panel shows the projection along the
 spectral axis of a portion of the FW extraction. The
 colours indicate individual clouds. The green (1), purple
 (2), yellow (3), pink (4), red (5), pink (6), cyan (7),
 and orange (8) fragments are identified as single clouds. This projection illustrates that, even after the removal of noise artefacts ("bad sources") the FW catalogue still contains a number of fragmented sources.}
\label{disconnected} 
\end{figure}

Figure \ref{efs} shows the FW and SCIMES extractions of \ce{^{13}CO} (3-2) emission in region 3 (see text and Fig.\,\ref{rcuts}) in the 59.72-km\,s$^{-1}$ velocity plane at 27.4-arcsec resolution. In the two panels, regions of space belonging to the cross-sections of different clouds are distinguished by different colours. The most prominent difference between the two extractions lies in the relative over-segmentation of the emission in the FW panel.  This is a known feature in FW extractions in which the watershed algorithm tends to break the emission into compact clumps that are accounted for as isolated features. A notable example is the large section of the SCIMES source extending from $34^{\circ}$ to $35^{\circ}$ of longitude in the mid panel of Fig.\,\ref{efs}. The selected velocity slice highlights how this extended SCIMES source becomes fragmented into adjacent clumps in the FW extraction. This behaviour is also observed in the example of segmentation of crowded and sparse fields provided in Fig. \ref{reg7} in Appendix \ref{exa}. 
In addition, as \cite{Rigby2019} points out, diffuse emission around the detection threshold can be identified as sets of disconnected voxels, clustered together as individual clumps (an example is given in Fig. \ref{disconnected}). These clouds are recognisable by their very irregular shapes and they were flagged as `bad sources' after a visual inspection in the FW catalogue \citep{Rigby2019}. These sources are excluded in the analysis that follows.

 Coherent sources at low SNR and areas of emission crossing the boundaries between tiles also belong to this category. These sources often present very irregular segmentation due to the difference in noise levels among tiles. Such discontinuities may also create small clumps that do not originate from features in the emission map but reflect changes in the emission in adjacent channels\footnote{With the FW parameterisation used for the segmentation of CHIMPS data, voxels with SNR $= 2$ can be included in a clump,
when they are directly connected to a clump with a peak SNR $> 5$ \citep{Rigby2019}.}. These inconsistencies are a consequence of performing the extraction on SNR maps. Such occurrences are, however, small in number and the total sample is only marginally impacted. 

The final catalogue published by \cite{Rigby2019} includes 4999 sources, $1335$ of which were classified as `bad sources' thought to arise from such artefacts.

If we directly compare the segmentation produced by FW with that of SCIMES on the same velocity plane (middle panel in Fig. \ref{efs}), the emission is 
segmented into fewer individual sources with SCIMES, generally covering larger areas than their FW counterparts. This characteristic of the SCIMES segmentation is supported by the analysis of the geometric and physical properties of its sources (see below), thus a cloud/clump is, in general, not characterised by a single maximum emission peak. SCIMES clusters consist of signals from different hierarchical levels of the emission dendrogram. The fragmentation induced by FW identifies pieces of the substructure as individual entities. In the framework of SCIMES, these clumps correspond to dendrogram branches. 
%

The introduction of artificial boundaries cutting through areas of less intense emission between peaks is a consequence of the watershed algorithm characterising disjoint clouds by single individual peaks. The volume and luminosity similarity criteria defining SCIMES clustering, instead,  allow for the grouping of emission from both the bright cores (i.e. dendrogram leave/peaks of the emission) together with their tenuous surrounding envelopes (bottom panel in Fig.\,\ref{efs}) into a single object, thus bypassing the impact of SNR discontinuities at the edges of adjacent tiles. 

\section{Physical properties}\label{basic_physical_properties}

\subsection{Cloud sizes}\label{cloud_size}

\begin{figure*}
\centering 
 	\includegraphics[width=1.0\textwidth]{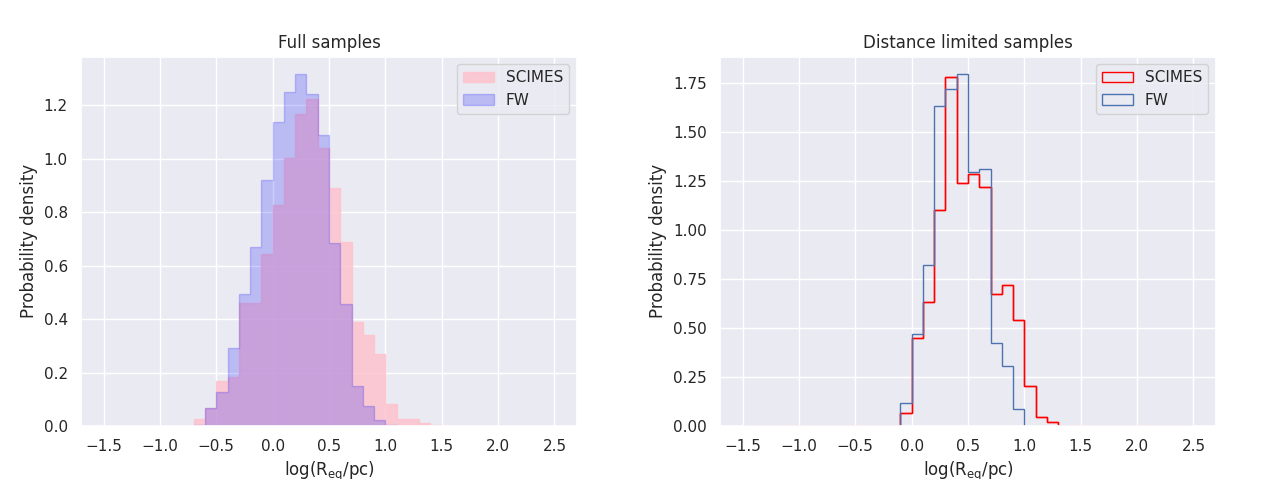} 
 	\caption{Distributions of equivalent radii in the SCIMES and FW segmentations (left panel). The right panels show the distributions for the distance-limited (8-12 kpc) samples.
  }
 	\label{new} 
 \end{figure*}

We estimate the size of the CHIMPS clouds by considering two 'approximate' radii associated 
with different characteristics of the emission. Adopting the definitions in \cite{Rigby2019}, we consider the equivalent radius $R_\mathrm{eq}$  as the radius of the circle whose area ($A$) is equivalent to the projected area of the source, 

\begin{equation}\label{req}
    R_\mathrm{eq} = d\sqrt{A/\pi}, 
\end{equation}

\noindent
where $d$ is the distance assigned to the source. The values of the equivalent radii associated with the SCIMES sources were calculated directly from the values of the exact areas produced by the {\sc Astrodendro} dendrogram statistics tools. 

For consistency in the comparison with physical properties defined in \cite{Rigby2019}, we also consider the geometric mean of the intensity-weighted rms deviations in the
$l$ and $b$ axes ($\sigma_l$ and $\sigma_b$), deconvolved by the telescope beam, and $d$ the assigned distance, 

\begin{equation}\label{rsigma}
    R_\sigma = d \sqrt{\sigma_l \sigma_b}.
\end{equation}

\noindent The "geometric radius" $R_\sigma$ provides a measure associated with the projected extent of the cloud in the $l$ and $b$ directions. Depending solely on the emission profile of the source,  $R_\sigma$ is less affected by the variations in the noise level in different areas of the survey (while $R_\mathrm{eq}$ has no dependence on the emission profile). $R_\sigma$ thus provides a more consistent measure than $R_\mathrm{eq}$ for the smallest and densest clumps where star formation is likely to be located (under the assumption that smaller clumps are centrally concentrated).


\begin{figure}
\centering 
 	\includegraphics[width=1
  \columnwidth]{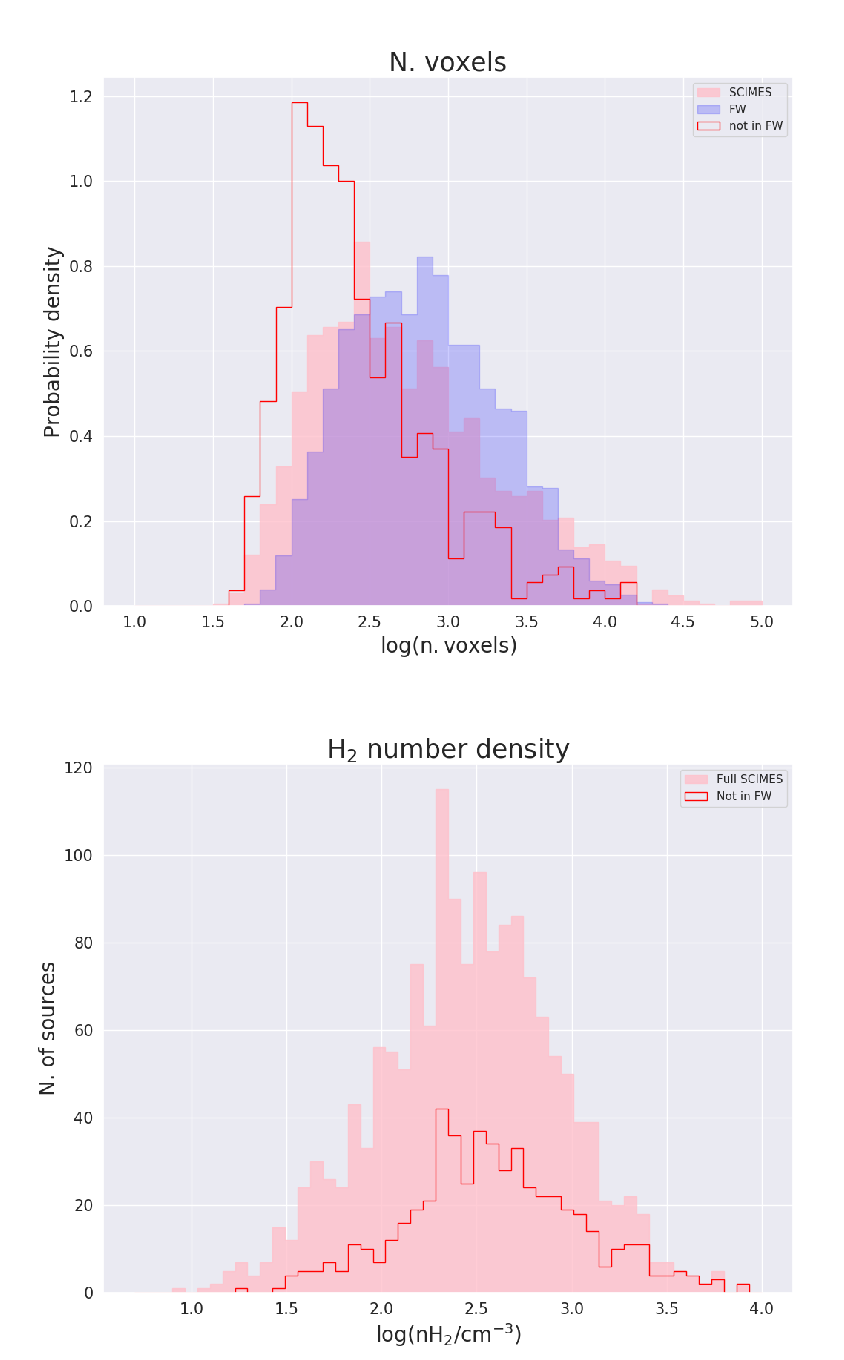} 
 	\caption{Distributions of numbers of voxels in the FW and SCIMES sources (top panel). The red outline histogram represents those SCIMES sources that do not contain any emission peak found in the FW catalogue. The bottom panel portrays the distributions of \ce{H_2} number densities for this subset and the whole SCIMES sample.
  }
 	\label{voxels} 
 \end{figure}

\begin{figure}
	\centering 
	\includegraphics[width=1.1\columnwidth]{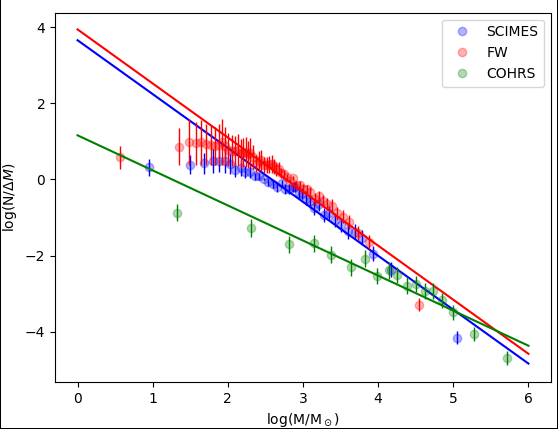} 
	\caption{Comparison between the data and the fitted functions for mass spectra. The dots indicate the centres of the mass bins. The colours refer to the method of extraction and survey.}
	\label{msi} 
\end{figure}

\begin{figure}
	\centering 
	\includegraphics[width=0.9\columnwidth]{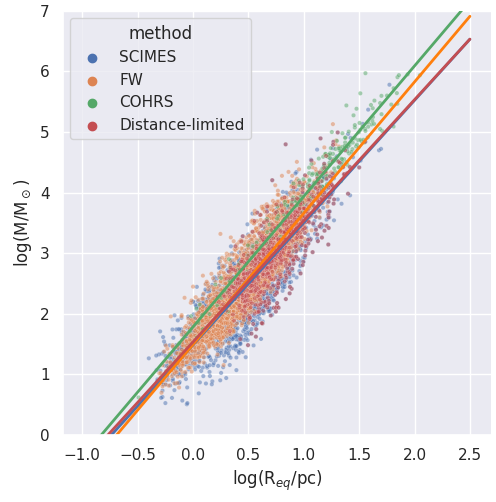} 
	\caption{Mass-radius relationship for CHIMPS and COHRS sources. Notice that the fit of the full SCIMES sample and the distance-limited subsample are nearly identical.}
	\label{mass_relations} 
\end{figure}

 We adopt a version of $R_\sigma$ scaled by a factor $\eta$ that considers an average emission profile. The constant $\eta$ is set to 2.  This value corresponds to the median value found by \citep{Rigby2019} for the FW extraction and it is a compromise between the commonly-used conversion $\eta = 1.9$ \citep{Solomon1987, Rosolwsky2006, Colombo2018} and $\eta = 2.1$, the median value we found using the alternative version of $R_\sigma$

\begin{equation}
R_\sigma = d \sqrt{\sigma_\mathrm{maj} \sigma_\mathrm{min}}, 
\end{equation}

\noindent
easily obtainable from the {\sc Astrodendro} statistical tools for finding the major and minor axes of the projected SCIMES sources.  The equivalent radius $R_\mathrm{eq}$ is used in all instances in which the radius enters the definition of a physical quantity. \cite{Rigby2019} also used the conversion factor $\eta$ in definitions where the comparison to different datasets required the use of $R_\sigma$.

A simple visual inspection of the segmented emission maps (see Fig. \ref{efs} for an example) reveals the over-segmentation produced by FW (more prominent in crowded fields). The high-value tail of the SCIMES distribution of $R_\mathrm{eq}$ in Fig. \ref{new}, relative to that from FW, confirms the higher number of larger clouds extracted by SCIMES. This result holds when heliocentric distances are constrained between $8$ and $12$ kpc. Following \cite{Rigby2019}, this specific distance-limited subsample is introduced as a `most reliable'  subsample against which we will compare any relationships between the physical quantities of the full sample to ensure that no bias is introduced by the choice of distance assignment. This set only includes 462 SCIMES sources with Galactocentric distances ranging from  4.0 to 8.5\,kpc. Within this distance range, the spatial resolution element between the nearest and most distant
sources differs by no more than 50\%, while the sub-sample covers a significant fraction of the full sample.

\begin{figure*}
	\centering 
	\includegraphics[width=0.87\textwidth]{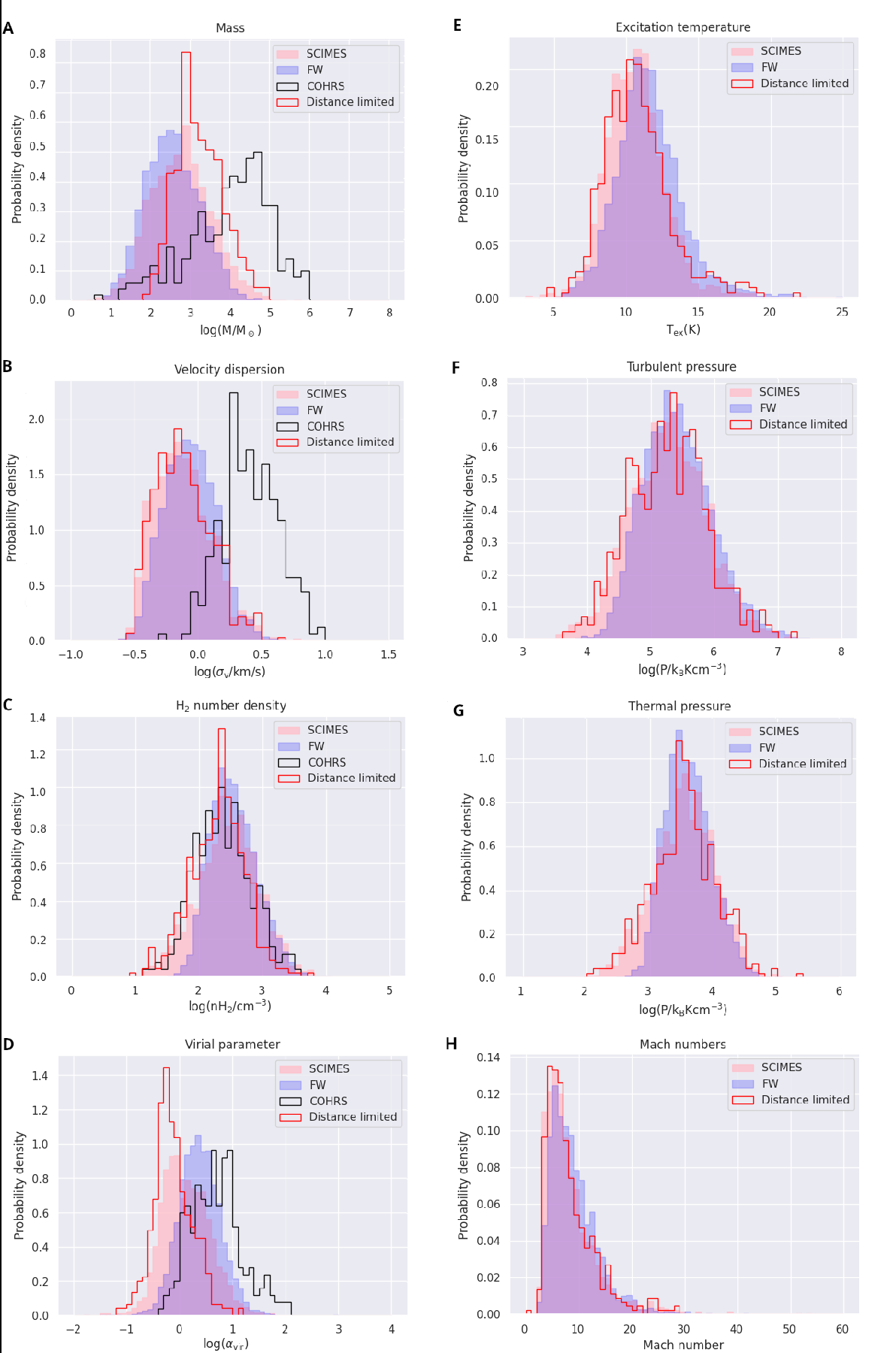}
	\caption{Panels A-H: distributions of total mass, equivalent radius, average number density, virial parameters, excitation temperature, turbulent and thermal pressure, and Mach numbers in the CHIMPS sources. Distributions of COHRS sources are also added for comparison when the data are available.
 }
	\label{histissimo}
\end{figure*}

In Fig.\ \ref{voxels}, we consider the source volumes, measured as the number of voxels that constitute each cloud as it is identified as an individual entity by the segmentation algorithms. 
We notice that, besides identifying large clouds in crowded fields (and thus being less prone to over-segmentation; see also Appendix \ref{exa}), SCIMES also extracts a significant number of smaller clouds, especially in sparse fields. The mean volume of all sources extracted amounts to 2191.8 voxels in SCIMES and 1307.1 voxels in FW. However, a dendrogram parameterization that matches the FW configuration described in \cite{Rigby2019} also produces 540 smaller clouds that do not contain any emission peaks arising from the FW extraction. These source may be found both in the proximity of similar emissions features identifid by the FW algorithm or in areas devoided of FW emission. This feature of the SCIMES segmentation becomes relevant in the calculation of velocity dispersions in sub-section \ref{velocity_dispersion}, in which the emission-weighted velocity channels spanned by a cloud are considered. Fig.\ \ref{voxy} in Appendix \ref{exa2} shows some examples of this set of sources. To ensure that these clouds are not low-emission artefacts constituted by low-density gas, we plot the distribution of their densities (Fig. \ref{voxels}), finding that it matches the distribution of the full SCIMES sample.

\subsection{Mass}\label{mass}

Once distances are assigned, the true size of each voxel in the SCIMES segmentation can be calculated. Its contained mass is then estimated through the column density cubes (see Section\,\ref{data}). The \ce{H_2} mass of the cloud is estimated by considering the mean mass per \ce{H_2} molecule, taken to be 2.72 times the mass of the proton, accounting for a helium fraction of 0.25 \citep{Allen1973}, and an abundance of $10^6$ \ce{H_2} molecules per \ce{^{13}CO} molecule \citep{Draine2011}.

The mass spectra for CHIMPS clouds and their fitted relations are displayed in Fig.\,\ref{msi}. The mass spectral indices found with a power law fit are $-1.41 \pm 0.05$ for SCIMES clouds, $-1.284\pm 0.02$ for FW and $-0.920\pm 0.04$ for the COHRS survey. The binning of the masses follows \cite{msi} with variable bin widths and fixed bin population of $2N^{2/5}$, with $N$ being the number of individuals in the entire population. This convention is adopted to remove biases due to binning and was previously used in \cite{Eden2015} and \cite{Eden2018}. The SCIMES index is consistent with $-1.6 \pm 0.2$ found by \cite{RomanDuval2010} and previous studies \citep{Sanders1985, Solomon1987, Williams1994}. FW is slightly below this value. COHRS masses are expressed in terms of the molecular gas luminosity and obtained by using the conversion factor $M = \alpha_\mathrm{CO} L_\mathrm{CO}$ , with
$\alpha_\mathrm{^{12}CO(1-0)} = 4.35\,\mathrm{  M}_\odot\, \mathrm{    pc}^{-2}\,\mathrm{km}^{-1}\mathrm{s}$, assuming a mean molecular weight of $2.8 m_\mathrm{H}$ per hydrogen molecule. The conversion factors were calibrated with
the \ce{^{12}CO}(1–0) assuming a line ratio $R_{31} =\ce{^{12}CO}(3-2)/\ce{^{12}CO}(1-0)$ to scale the calculated properties directly
to physical properties \citep{Colombo2018}. The COHRS sample shows the greatest discrepancy, hinting that a single power law might not be applicable to all tracers of the same molecular clouds. The slope of the COHRS spectrum produces the best fit for values around $M>10^5 M_\odot$ (where it becomes similar to the fits of the SCIMES and FW samples). The flatter slope in \ce{^{12}CO} suggests that this SCIMES segmentation detects fewer small individual leaves that we could extract with \ce{^{13}CO} because they get grouped into larger structures connected by more diffuse material (and, perhaps, a number does not even appear as peaks in the \ce{^{12}CO} emission because of the gas being optically thick). In this scenario only the diffuse gas around the clumps is detected, suggesting that the COHRS sample, identifying more massive structures, is incomplete at smaller masses (see discussion below).

The turnover at $\sim 300 M_\odot$ is an indicator 
of the completeness limit of the data.  This is the mass limit below which sources are not dependably extracted and therefore their distribution cannot be fitted by any power law. This limit depends on the size in both spatial and spectral axes, the
local noise level, and the source density profile in addition to the
total mass. \cite{Rigby2019} show that there is no single completeness limit in the CHIMPS data as the completeness limit is distance-dependent.



Vital to an accurate mass estimation is a precise distance assignment. The typical uncertainty on the distances estimated from the Bayesian
distance algorithm is $\sim 0.3$ kpc \citep{Reid2016}, which affects shorter distances the most (30\% at 1\,kpc) but falls to a few per cent already at 5\,kpc. Taking into account the error on the conversion CO-to-\ce{H_2} conversion factor and column density estimation \citep{Urquhart2018, Rigby2019}, we estimate a typical error in cloud mass of order 30-40 per cent. In addition, the distance assignment (as well as all other calculated parameters) is very likely to be contaminated by uncertainties in the assumptions and approximations in the variety of methods considered in the various surveys. Section \ref{note} presents a comparison between mass distribution derived from random distance assignments, suggesting distance assignments make no significant difference to the full-sample statistics.

\begin{figure*}
\centering
	\includegraphics[width=1\textwidth]{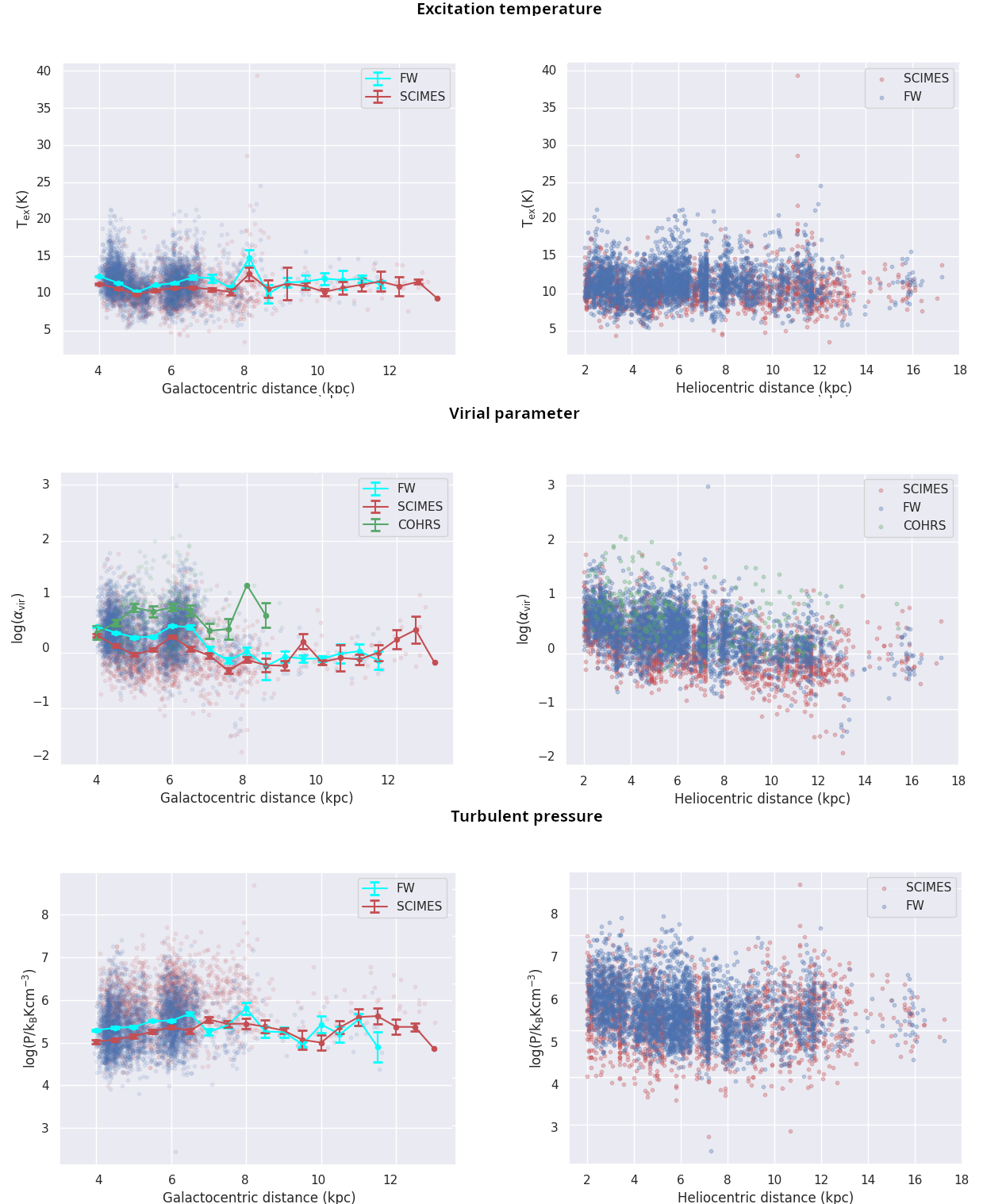}
	\caption{Various properties measured for the CHIMPS and COHRS (when data are available),  namely the equivalent radius, mass, mean number density, excitation temperature, virial parameter and turbulent pressure as functions of both Galactocentric (left) and heliocentric (right) distance. For Galactocentric distances, we have plotted trendlines and error bars. The trendlines connect the mean values of $0.5$ kpc wide bins. The error bars are the standard errors of the means. The rise in the density plots at low heliocentric distances may be considered an indicator of a resolution bias. This bias is visible in the distribution of $R_\mathrm{\rm req}$ with heliocentric distance. }
	\label{ultrissimo}
\end{figure*}

Fig.\,\ref{mass_relations} shows the mass-radius relationship for sources in CHIMPS extracted with both the FW and SCIMES methods. Power law fitting produces slopes of $2.02 \pm 0.02$  and $1.97 \pm 0.02$ for FW and SCIMES respectively. 
The distance-limited sample is fitted with $ 1.93 \pm 0.06$. The values are similar to the power law exponent of  $2.36$ found for molecular clouds in the GRS \citep{RomanDuval2010}   
The scatter in the CHIMPS data is much larger than that in the GRS \citep{Rigby2019} and probably relates to the large difference in resolution, and it is comparable to the scatter in the ATLASGAL data, which were extracted at similar resolution ($\sim 20$\,arcsec). Dense clumps in ATLASGAL are found to follow a shallower power law with exponent $1.65$ \citep{Urquhart2018}. 
COHRS sources ($2.15 \pm 0.03$) have been added for comparison. As expected, the larger structures detected through \ce{^{12}CO} emission result in the larger masses in panel A of Fig.\,\ref{histissimo}, and the distributions with distance in Fig.\,\ref{ultrissimo}. CHIMPS and COHRS trendlines also follow a similar pattern, suggesting that the segmentation of COHRS identifies the more extended counterparts of CHIMPS objects.  

Panel A in Fig.\,\ref{histissimo}  compares the distributions of mass in the two CHIMPS emission extractions with that mass in COHRS clouds. The calculation for mass estimation from CO luminosities in COHRS is described in \cite{Colombo2018}. The mass distribution reflects the size distribution of the clouds for the SCIMES and FW segmentations. 

The mass distribution as a function of heliocentric and Galactocentric distances of the full sample is presented in Fig.\,\ref{ultrissimo}, where the trend at small Galactocentric distances is likely to be an artefact originating from the small number of sources in the initial bin (3.5-4.0\,kpc) and the position of the centre of the bin in the plot. 

\begin{figure*}
\centering
	\includegraphics[width=1.1\textwidth]{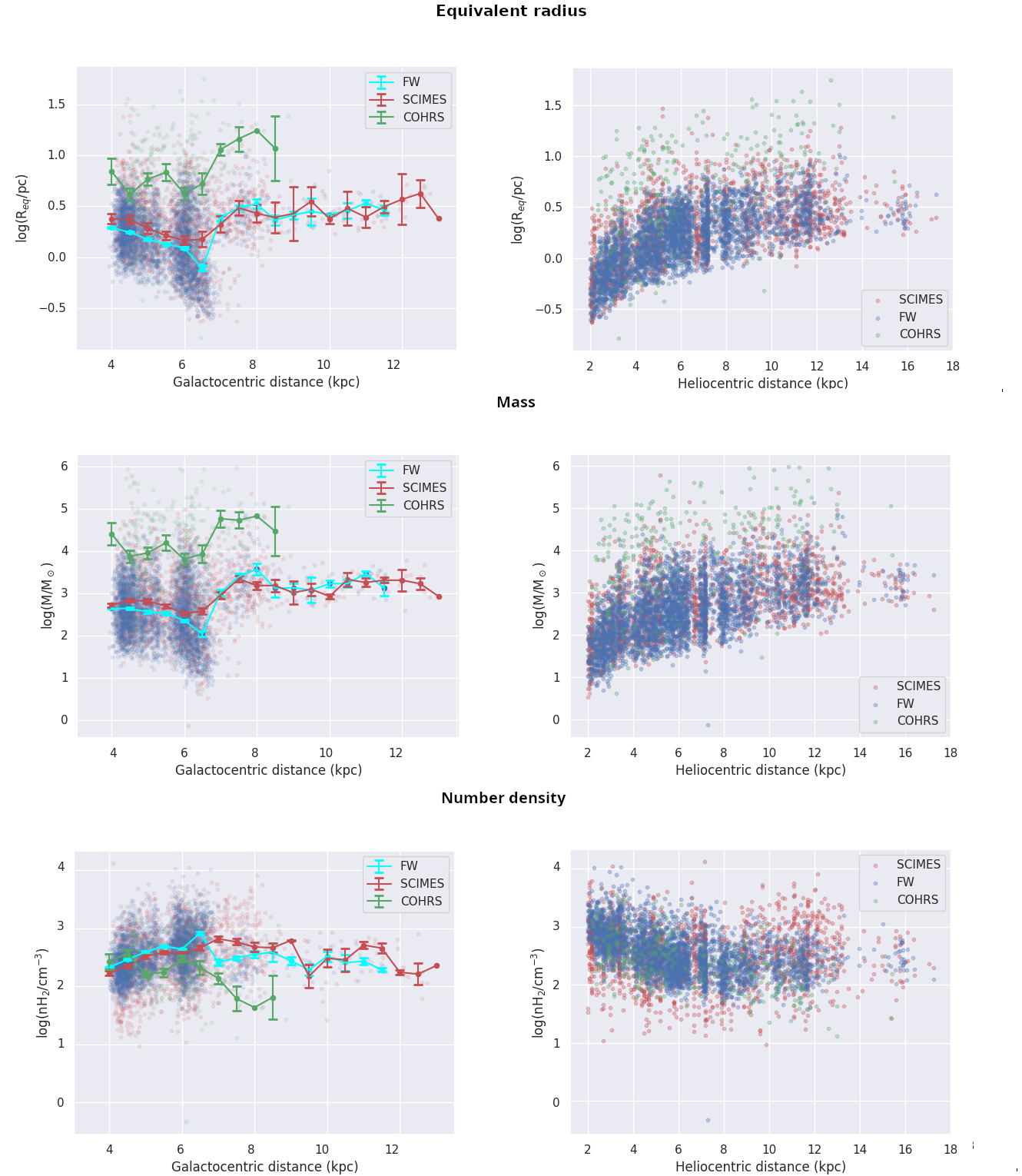}

\end{figure*}

\subsection{Hydrogen number density}\label{hydrogen_number_density}

The mean (volumetric) particle density (or number density)
over the approximate volume of a cloud (assuming 2D to 3D symmetry) is calculated as

\begin{equation}\label{nH2}
    \overline{n}(\mathrm{H}_2) = \frac{3}{4\pi}\frac{M}{\mu m_p R_{eq}^3}, 
\end{equation}

\noindent
where $M$ is the mass of the cloud, $ \mu \mathrm{m}_p$ ($= 2.72 \mathrm{m}_p$)  is the mean molecular weight. 

The distribution of molecular hydrogen number densities extracted from CHIMPS via FW and from CHIMPS and COHRS by SCIMES is reported in panel C of Fig.\,\ref{histissimo}. The larger masses and greater radii found in COHRS clouds result in a distribution of mean molecular hydrogen density that is comparable to the ones obtained for the SCIMES and FW segmentations.


We notice that the distributions of \ce{H_2} number densities exhibit values much less than the critical density of the \ce{^{13}CO} (J=3--2) transition. In a clumpy medium, the average density may be an underestimate of the typical density at which most emission originates and the \ce{H_2} number density assigned to each cloud represents the average density over the entire (approximated) volume of the cloud. This average value accounts for both clumps with a density over the critical threshold and areas of far more rarefied gas.

Gas with densities lower than the critical density will also be warmer than the calculated excitation temperature \citep{Rigby2019}. However, it may still emit in a sub-thermal mode in which the energy level populations are not distributed according to the Boltzmann distribution.
This underestimate in the gas temperature is mirrored in overestimates
in the gas column density \citep{Rigby2019}. The distribution of mean excitation temperatures of the FW extraction of CHIMPS clouds is found to have a mean value of 11.5 K, which matches the expectation for molecular structures covering the size regime from cores, through clumps, to clouds \citep{Bergin2007}. Sub-thermal emission can therefore be assumed not to be a dominant effect in the \ce{^{13}CO} emission \cite[see also][]{Rigby2019}. 

The unexpected left tail in the distribution of SCIMES mean number densities
should not necessarily be considered an indication of smaller volumes or masses in disagreement with our previous results, but rather arising from the inaccurate spherical approximation of larger irregularly-shaped clouds. The approximation is aggravated by using the equivalent radius to match the cloud's extension both along the Galactic coordinates and the line of sight. 
This is particularly evident for clouds with large aspect ratios (filamentary) are more likely to have a "depth" which is similar to the smaller dimension of the projected cloud (i.e. the width of the filament). In which case, $R_\mathrm{eq}$ estimated from the equivalent area will provide an overestimation of depth, and consequently an underestimation of the cloud's density.

\subsection{Velocity dispersion}\label{velocity_dispersion}

The velocity dispersion ($\sigma_v$) measures the statistical dispersion of velocities about the mean velocity for a molecular cloud. In the clump-finding implementation of FW provided in the JCMT Starlink software suit, $\sigma_v$ is estimated as the RMS deviation of the velocity of each voxel centre from the clump velocity centroid \citep{FW}. The FW catalogue adopts this as the measure of the extent of a cloud along the spectral axis. For a cloud with a Gaussian distribution of velocities, this definition of $\sigma_v$ corresponds to the standard deviation of the distribution with mean value at the centroid velocity. Equivalently, SCIMES derives its velocity dispersion from the intensity-weighted second moment of velocity through the {\sc Astrodendro ppvstatitics} function. The distributions of the velocity dispersion in Fig.\ \ref{histissimo} reflect the difference in size of the clouds extracted by the two methods, these being related via Larson's relations.  Although SCIMES tends to extract overall bigger sources, this extraction also exhibits a significant number of smaller clouds that are not matched to FW emission (see Section \ref{cloud_size}). This subset contributes systematically smaller values of  $\sigma_v$, shifting the overall distribution, which then acquires a lower mean (0.89 km\,s$^{-1}$) than FW (0.98 km\,s$^{-1}$).

\begin{figure*}
	\centering 
\includegraphics[width=2.17\columnwidth]{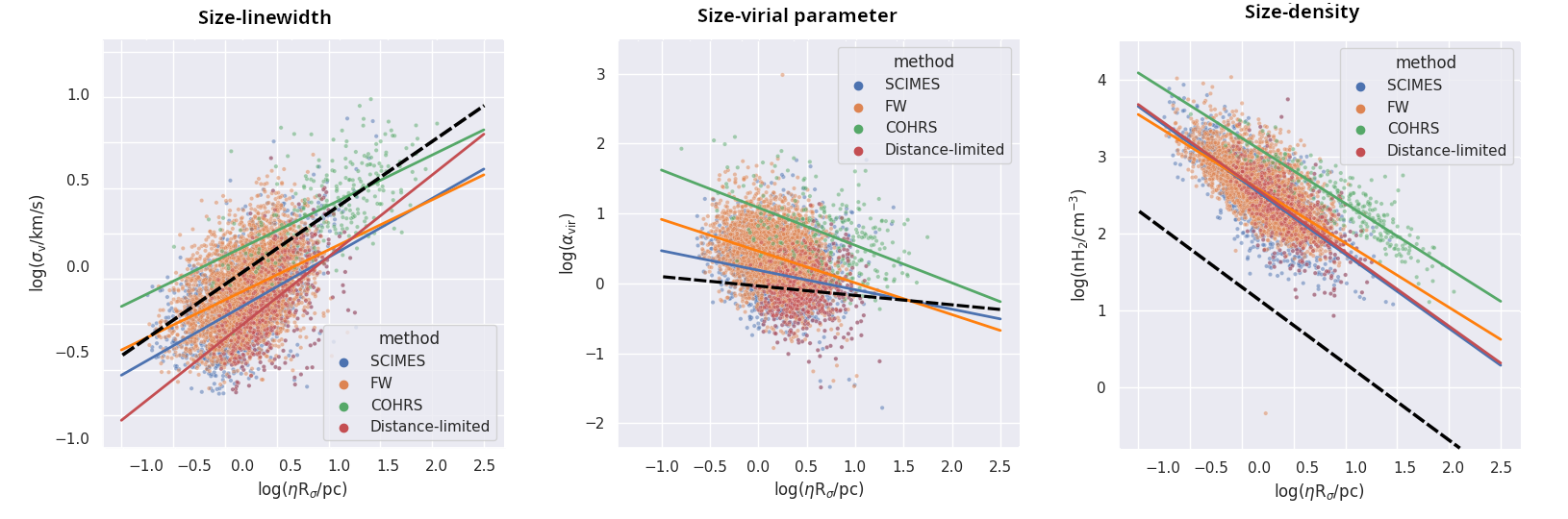} 
	\caption{Size-linewidth (right panel), size-virial parameter (central panel), and
size-density (left panel) relationships for the CHIMPS and COHRS
sources. The size parameter is the scaled intensity-weighted rms
size, $\eta R_\sigma$, for which $\eta = 2.0$. Fitting lines are shown where
a correlation is found between the quantities considered. The dashed lines indicate the Larson relations.}

\label{scaling_relations_f}
\end{figure*}

For the COHRS plots, we consider the non-extrapolated and non-convolved data, see \citep{Colombo2018}.
In general, the larger the size of a cloud, the wider the distribution of velocities of its particles, thus its velocity dispersion. The velocity dispersion causes the broadening of linewidths in CO observations. This fact is mirrored in the distribution of velocity dispersions in the clouds of the COHRS catalogue and their size-linewidth relation in Fig.\,\ref{scaling_relations_f}. Line widths are expected to be larger in \ce{^{12}CO} because of the high optical depths suppressing the peak intensities as well as tracing larger structures with larger turbulent velocities.


\subsection{The virial parameter}\label{the_virial_parameter}

The virial parameter encodes the dynamic state of a molecular cloud, assuming that the cloud is capable of sustaining virial equilibrium.


The virial parameter is defined as the ratio of a cloud's spherically symmetric virial mass to its total mass ($M$)

\begin{equation}\label{alpha_vir}
    \alpha_\mathrm{vir} = \frac{3\sigma_v^2\eta R_\sigma}{GM}
\end{equation}

\noindent
where $G$ is the gravitational constant. 
This definition \citep{Rigby2019} assumes a radial density distribution 
$\rho(r) \propto r^{-2}$ \citep{Maclaren1988} 
and includes 
$R_\sigma$ to account for the median emission profile. The intensity-weighted radius 
reinforces the gravitational energy in those regions where the density is higher. 

Approximating a source as a spherically symmetric distribution of density introduces a factor-of-two uncertainty in the estimation of the virial parameter. This arises from both characterising the source by a single radius and from choosing this particular radial profile. This error will be systematic to a large extent, and likely to affect both segmentations in the same fashion.

In the absence of a strong magnetic field or external pressure, $\alpha_\mathrm{vir}$ equals $1$ when the clouds are in virial equilibrium. A value $\alpha_\mathrm{vir} = 2$ indicates that the gravitational energy equals the kinetic energy in the cloud. Values of  $\alpha_\mathrm{vir}$ smaller than $1$ characterise an unstable, collapsing system (when other sources of supporting pressure are absent). A dissipating system, dominated by kinetic energy, is characterised by $\alpha_\mathrm{vir} > 2$. While $1<\alpha_\mathrm{vir} <2$ indicates approximate equilibrium. These clouds may be free-falling and small values of the virial parameter may indicate other support or observation biases \citep{Traficante2018}. It has been suggested that the heightened velocity dispersions due to rapidly infalling gas in collapsing cloud fragments may still raise the cloud's value of the virial parameter to $\sim 2$ \citep{Kauffmann2013}. This would be the case of the smaller FW clouds, identified around single high-emission, high-density peaks. Fragments with $\alpha_\mathrm{vir} \ll 2$ are more likely to host and be supported by strong magnetic fields or to house ongoing high-mass star formation. In the absence of these conditions, their life would be too short to allow for their detection  \citep{Kauffmann2013}.

The distribution of the virial parameter in CHIMPS and COHRS  is presented in panel D of Fig.\,\ref{histissimo}. The SCIMES distribution indicates that a large number of clouds in this segmentation are gravitationally unstable or in approximate equilibrium. 

Fig.\,\ref{ultrissimo} shows the virial parameter as a function of the Heliocentric and Galactocentric distances, respectively. A closer look at the trendlines in Fig.\,\ref{ultrissimo} reveals a hint of a slightly increased $\alpha_\mathrm{vir}$ inside 7\,kpc, or perhaps in the spiral arms. This trend may be due to the errors on the means of the bins increasing significantly at large radii. The decrease of the virial parameter as a function of heliocentric distance reflects the mass trend shown in 
Fig. \ref{ultrissimo}. We notice that this feature was also found in the SEDIGISM survey \citep{Schuller2017} and may thus be an indication of some observational bias.

\subsection{Scaling relations}\label{scaling_relations}

To continue the comparison with the analysis proposed in \cite{Rigby2019} for the FW sample, we now consider the scaling relations between molecular-cloud properties. Applying a power-law fit to the size-density relation shown in Fig.\,\ref{scaling_relations_f} produces average number densities proportional to $R^a$ with $a = -1.01 \pm 0.02$ for SCIMES clouds ($a = -0.97 \pm 0.02$ for the distance-limited subsample, and $a =-0.99\pm 0.05$ in the FW case). For COHRS clouds $a$ equals $ -0.85\pm 0.03$. 
The fits of FW and SCIMES sources both produce values of $a$ similar to the original scaling relation $a = -1.1 \pm 0.05$ found by \cite{Larson1981}.

 A fit to the size-velocity dispersion relation produces $\sigma_v \propto R^a$ with $a=0.31\pm 0.01$ for SCIMES clouds ($a=0.42\pm 0.03$ for the distance-limited subsample), and $a = 0.34\pm 0.01$ in the FW case). Both values are similar to the original scaling relation $a = 0.38 \pm 0.14$ found by \cite{Larson1981}  over a factor of 30 in size, which was originally interpreted as evidence that the internal motions of molecular clouds follow a continuum of turbulent flow inherited from the ISM at larger scales. For the COHRS clouds $a =0.28\pm 0.02$. 

SCIMES clouds that are characterised by smaller values of the virial parameter ($< 0.6$) fall in a size range between 2 and 20 pc. These clouds include the smallest, most compact sources, and the most likely sites of star formation. The size-virial parameter relation fit produces $\alpha_v \propto R^a$ with $a=-0.25 \pm 0.03$ for SCIMES clouds. The distance-limited subsample provided the least precise fit with $a=0.09 \pm 0.06$, consistent with zero. The discrepancy between the values for the full sample and the distance-limited sources is due to the lack of statistical correlation (Spearman correlation coefficient $ = 0.01 $  with p-value $= 0.03$) between $R_\mathrm{eq}$ and $\alpha_\mathrm{vir}$ in the reduced set, in addition to the large error produced by the chi-square fitting algorithm. Fitting the FW sources yields $a=-0.45 \pm 0.02$. The slopes found above for the SCIMES and FW sources are significantly steeper than the original scaling relation $a=-0.14$ found by \cite{Larson1981}. The discrepancy may be due to the varying mass completeness as a function of distance. A factor $a=-0.54 \pm 0.06$  was found for COHRS clouds.


\subsection{Free-fall and crossing times}\label{freefall_and_crossing_times}

The free-fall timescale, $t_\mathrm{ff}$, represents the characteristic time that would take a body to collapse under its own gravitational attraction. As mentioned above, $t_\mathrm{ff}$ depends solely on the density and the densities of the chemical species of the gas. In terms of the molecular hydrogen mean number density discussed in the previous sub-section, 

\begin{equation}\label{free_fall}
    t_\mathrm{ff} =\sqrt{\frac{3\pi}{32G\mu m_{p}\overline{n}(H_2)}}.
\end{equation}

The crossing timescale, $t_\mathrm{cross}$, corresponds to the time it takes a disturbance to cross the system at the sound/signal speed in the medium. The length of 
$t_\mathrm{cross}$ is directly proportional to the size of the system and inversely proportional to the velocity dispersion= of the gas:

\begin{equation}
  t_\mathrm{cross}  = \frac{2 R_\mathrm{eq}}{\sigma_v}. 
\end{equation}

\begin{figure}
\centering	\includegraphics[width=\columnwidth]{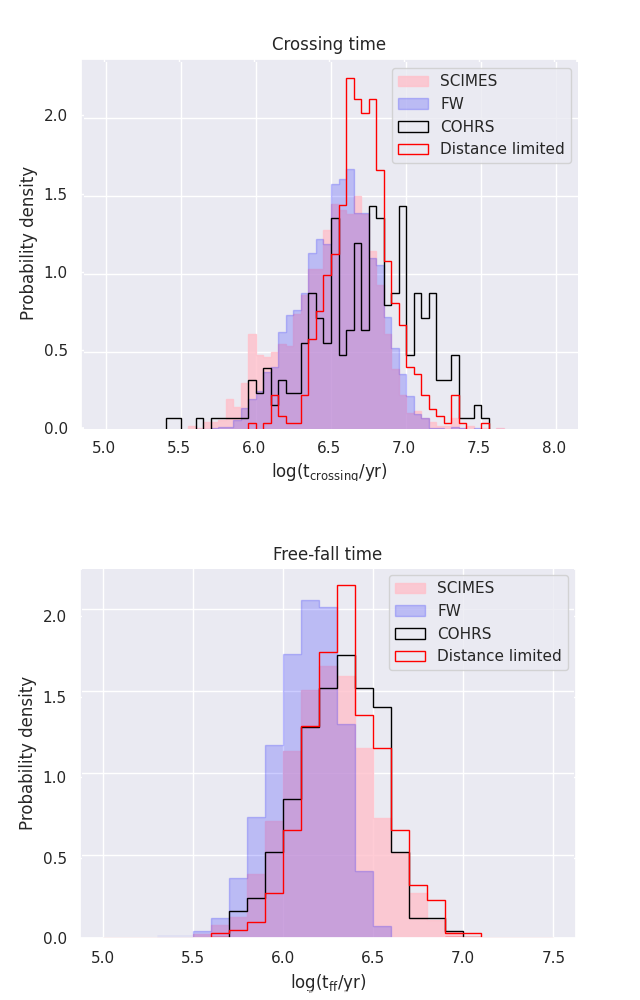}
\caption{Distributions of the  crossing and free fall
timescales associated with the CHIMPS \ce{^{13}CO} (3 - 2)
sources in the FW (blue) and SCIMES (red), and COHRS
(black) catalogues.}
\label{exia_times} 
\end{figure}

The distributions of these timescales for the two segmentations of CHIMPS and COHRS are compared in Fig.\,\ref{exia_times}. FW and SCIMES crossing times present similar distributions.

The left and right tails in the distribution of crossing times in SCIMES reflect the corresponding distribution of velocity dispersions. The distribution of free-fall timescales evidences the lower average surface densities of the larger SCIMES clouds, and may also originate from the elongated clouds where the volume density was underestimated by the spherical approximation with radius $R_\mathrm{eq}$.

\subsection{Excitation temperature}\label{excitation_temperature}

Excitation temperatures are assigned to clouds by considering the mean temperature contained within the cloud assignments in the maps constructed in Section\,\ref{data}. The distributions of excitation temperature in the FW and SCIMES segmentations of the \ce{^{13}CO} (3-2) emission in CHIMPS are shown in panel E of Fig. \ref{histissimo}. The temperatures from the SCIMES catalogue are systematically lower than FW temperatures. The average SCIMES excitation temperature is $10.19 \pm 0.040$\,K while FW clouds have a mean of $11.54 \pm 0.039$\,K. Although SCIMES detects, in general,  more diffuse and thus potentially warmer material, the higher average temperature estimated in the FW sample is likely to be due to the SCIMES clouds being larger and thus extending to lower CO brightnesses, which results in a lower inferred excitation temperature when the beam filling is assumed to be $\sim 1$.
CHIMPS excitation temperatures do not vary significantly with distance (Fig.\,\ref{ultrissimo}). As a function of the Galactocentric distance, the two segmentations show no obvious (difference in) biases and no overall gradient of the excitation temperature (the initial decreasing gradient with Galactocentric distance cannot be confirmed due to the lack of information at distances shorter than $3.5$ kpc from the Galactic centre in CHIMPS). This contrasts with the probable gradient in the interstellar radiation field \citep{Maciel2007}, dominated by cosmic-ray heating or (less likely) by internal heating. The density regime probed by CHIMPS, however, provides enough shielding to contrast this effect.

Arm radii ($\sim 4.5$, $\sim 6.5$, and $\sim 7.5$ kpc, see section \ref{distance_assignments}
) only see an increase in source counts, which increases the detected wings of the scatter distribution to higher $T_\mathrm{ex}$, but does not result in a significant change in the mean. 

The high-temperature outliers in the SCIMES segmentation have coordinates and distances corresponding to those of the star-forming region W49 ($l \approx 43.2^\circ $, $b \approx 0.0^\circ$ at 11.1\,kpc). This region is considered
extreme as it has column densities dust temperatures, and luminosity per unit mass \citep{Nagy2015} consistent with those found in
luminous and ultraluminous infrared galaxies \citep{Solomon1997, Nagy2012}. The region also has an overabundance of ultracompact HII regions \citep{Urquhart2013}.

\subsection{Turbulent pressure}\label{pressure}

The three-dimensional velocity dispersion ($3\sigma_v^2$) can be decomposed into its thermal 

\begin{equation}\label{themal}
\sigma^2_{\rm T} = k_{\rm B} T_\mathrm{ex}/ \mu m_{\rm p}
\end{equation}

\noindent
and non-thermal (turbulent) 

\begin{equation}\label{non_themal}
    \sigma_\mathrm{NT}^2 = 3\sigma_v^2 - \sigma^2_{\rm T}
\end{equation}

\noindent
components, where the one-dimensional velocity dispersion is defined in sub-section \ref{velocity_dispersion}. 

The turbulent pressure is then defined as

\begin{equation}\label{tp}
    P_\mathrm{turb}/k_{\rm B} = \mu m_{\rm p} \, \overline{n}(H_2)\,\sigma_\mathrm{NT}^2/k_{\rm B}
    \quad \mathrm{K\,cm}^{-3}, 
\end{equation}


This is the internal pressure of the clouds arising from the turbulent motions of molecular gas. The turbulent pressure distributions in panel F of Fig.\ref{histissimo} show that SCIMES sources tend to have lower pressure than their FW counterparts. The lower pressures appearing in the SCIMES distribution are likely to be a consequence of 
SCIMES' smaller velocity dispersions $\sigma_v$ entering definition \ref{tp} since at larger scales we would expect clouds to manifest higher turbulent pressure.
We notice that the three peaks characterising the distribution of turbulent pressures in the distance-limited sample are likely to arise from the difference in the environmental density of the sources located within spiral arms \citep{Bonnell2006}.
The median values of the two distributions are comparable with SCIMES having a median of $2.5 \times 10^5\,\mathrm{K\,cm}^{-3}$ and FW of $4\times 10^5$ $\mathrm{K\,cm}^{-3}$. Both these values agree with the total mid-plane pressure in the Solar neighbourhood ($\sim 10^5\,\mathrm{K\,cm}^{-3}$).


The distribution of $P_\mathrm{turb}/k_{\rm B}$ with helio- and Galactocentric distance are given in Fig.\,\ref{ultrissimo}, respectively. The range of $P_\mathrm{turb}/k_{\rm B}$ covered by both distributions is consistent with the mid-plane values \citep{Rathborne2014}.

The thermal pressure can be defined as

\begin{equation}
    P_\mathrm{thermal} = \overline{n({\rm H}_2)} \,k_{\rm B}\, T_\mathrm{ex}.
\end{equation}

Thermal pressure distributions are presented in panel G of Fig.\,\ref{histissimo}. The turbulent pressures are found to be $\sim 60$ times greater than the corresponding thermal pressures. Lower average densities result in lower pressures associated with the COHRS sample.

\subsection{Mach numbers}\label{mach_numbers}



Panel H of Fig. \ref{histissimo} represents the distributions of Mach numbers $\mathcal{M} = \sigma_\mathrm{NT} / \sigma_{\rm T}$ of the sources in the FW and SCIMES segmentations. The distributions look similar, both peaking in the supersonic regime ($\mathcal{M} \sim 5$) and extending out to higher Mach numbers. 

The difference in the distributions vanishes as the tails of the distributions flatten out past $\mathcal{M}=20$ where fewer large enough clouds to sustain these hypersonic regimes are found.

\section{Conclusions}\label{Conclusions}

This article presents a cross-correlation of the properties of individual clouds in two different segmentations of the \ce{^{13}CO} ($3-2$) emission in the CHIMPS survey: one obtained with the watershed algorithm {\sc FellWalker} and the other with the dendrogram-based SCIMES. These methodologies yield different numbers of molecular clouds (1586 with SCIMES while FW yields a reliable set of 3665 sources) but produce largely consistent results with similar ranges in masses, equivalent radii, mean number densities, and velocity dispersions. 
The distributions of mean number densities, masses, virial parameters, and dynamic timescales all reflect the differences in volumes and geometries found in the two segmentations.
A word of warning should however be spent on the cross-correlation of the physical properties of individual sources between the two catalogues. Different definitions, numerical implementations, and design choices within each method influence the estimated value of a given physical quantity and those derived from it. 
Additionally, the SCIMES extraction of \ce{^{12}CO} ($3 - 2$) in COHRS is considered as a term of comparison with a different tracer over the same area spanned by CHIMPS. 
This particular transition of the \ce{^{12}CO} isotopologue is, in general, a more optically thick tracer than \ce{^{13}CO} ($3 - 2$). In practice, this implies that the COHRS segmentation traces lower-density regions of the molecular clouds, that are not detected in CHIMPS.  The line-widths for the COHRS clouds will thus be naturally wider than those found through both SCIMES and FW (Section\,\ref{velocity_dispersion}). Probing lower-density emission, COHRS detects larger structures than CHIMPS. 
To a lesser degree, the inconsistent results in the SCIMES segmentations of \ce{^{12}CO} and \ce{^{13}CO} emission 
can also be traced back to the 
different SCIMES parameterisations chosen for the segmentations in \cite{scimes}.  Since the optimum parameter values are determined, to a large extent, by the characteristics of the data, these two effects are closely related.


A closer look at the distribution of the assigned SCIMES heliocentric distances (Fig.\,\ref{dist}) and the independently generated Galactocentric distances reveals that both distributions display the same features as the FW assignments. The difference in distance assignment has supposedly little influence on the distance-dependent physical properties. Size-linewidth, size-density (Fig.\,\ref{scaling_relations_f}) and size-virial parameter plots for the CHIMPS clouds, also reveal similar relations. An identical situation is reported by \citep{Lada2020} in their studies of mass-size relations \citep{Larson1981} and the GMC surface densities in Galactic clouds. \cite{Lada2020} compared data from the SCIMES \citep{Rice2016} and FW \citep{MD2017} extractions of \ce{^{12}CO} in the low-resolution CfA-Chile survey \citep{Dame2001}. The mass-size relation they found did not appear to be particularly sensitive to differences in the two methodologies used for the emission segmentation. 


Although the two segmentation methods produce similar statistical results when applied to the full survey with the chosen parameterisation, on the scale of individual clouds the situation may differ. The SCIMES extraction (subsamples with SNR > 10) includes larger sources than FW both in crowded and sparse environments. Notice that the full SCIMES catalogue also includes a significant number of smaller sources, most likely found in sparse fields, these clouds have no FW counterparts, see sub-section \ref{cloud_size}. This feature underscores the difference in the paradigms that characterise the two methods and the 
difficulty of establishing a one-to-one correspondence across catalogues produced by different algorithms.
In crowded fields such as large star-formation complexes like W\,43 ($l = 30\fdg 8$, $b = 0\fdg 0$), FW tends to split clouds into smaller clumps. Visual inspection reveals that the FW clumps have touching sharp borders (see Fig.\,\ref{efs}) whereas SCIMES identifies a single structure (this is also evident in the cross-sections of the clouds in the velocity plane shown in Fig. \ref{efs}). The introduction of artificial boundaries between emission peaks is a consequence of the watershed algorithm which characterises disjoint clouds by single individual peaks. This method cuts the valleys between peaks into separate assignments, thus splitting the envelopes of more rarefied structures enclosing denser clumps. This defining characteristic makes FW and similar methods better suited to extract sources in less crowded fields or to identify compact cores in crowded fields through a careful selection of the configuration parameters. With the chosen parameterisation, SCIMES, on the other hand, registers such structures as part of a single entity, thus proving to be more sensitive to tenuous emission in complex gas distributions and crowded fields.
This suggests that on individual clouds the application of FW and SCIMES with suitable parameterisations may serve different purposes: the extraction of dense clumps and cores for the former and the identification of full outer envelopes and cloud contour for the latter. The efficiency of a segmentation algorithm is thus strictly linked to the task for which it is used. The dendrogram approach adopted in SCIMES offers the additional advantage of recording information on both the full hierarchy of emission (trunk) and the individual peaks (leaves) which can be combined with the analysis of clusters at a specified spatial scale.

A number of cases of disconnected FW sources are present in all CHIMPS regions. Fig.\,\ref{disconnected} shows some examples of these fragmented sources that are left in the FW catalogue even after the removal of the noise artefacts \citep{Rigby2019}. This feature is not present in the SCIMES extraction and it is thought to arise from the implementation of the FW method. Establishing an accurate relationship between the results of
the FW and SCIMES on the scale of individual clouds would however require the accurate analysis of substructures in individual clouds in different environments. This would allow for the
identification of FW clouds within the SCIMES dendrograms,
matching them with branches and sub-branches.



By the definition of emission dendrogram, the extraction produced by SCIMES is more sensitive to the overall gas distribution in the region in which the segmentation is performed. For $^{12}$CO COHRS, SCIMES produces structures that are $\sim$100 times larger in $(l,b,v)$ volume. This is unsurprising, because of the high optical depth, self-absorption and lower critical density associated with $^{12}$CO.

The mass spectrum in  Fig.\,\ref{msi} shows that  SCIMES and FW $^{13}$CO power-law fit gradients are similar, however, the COHRS spectrum is flatter than either. SCIMES and FW mass distributions are similar for $^{13}$CO. COHRS $^{12}$CO clouds, on the other hand, are two orders of magnitude more massive. The mass-radius relation is similar for the three samples considered. The linear sizes of the FW and SCIMES sources are similar, but clouds in the distance-limited SCIMES sample have an overall larger size.  COHRS clouds are $\sim$10 times larger than the $^{13}$CO sources. 

When mean excitation temperatures are considered, the SCIMES sources present lower temperatures than those found in the FW extraction. Turbulent pressure reduces in SCIMES. A similar behaviour is observed for thermal pressure, but its distribution presents a tail to lower pressures in SCIMES clouds. COHRS clouds have $\sim$10 times lower pressures.

The distribution of volume densities is similar for all extractions and species, but the COHRS distribution presents a lower tail. SCIMES clouds have smaller virial parameters than the FW ones. The value of $\alpha_\textrm{vir}$ in COHRS values are $\sim$10 times higher. SCIMES and FW sources also display similar Mach numbers with COHRS being slightly lower.

Our comparison thus suggests that there are some systematic differences in the physical parameters of clouds that result from the extraction method. The very large differences between the $^{12}$CO and the $^{13}$CO samples are due to high optical depth and lower critical density in the former.

\section{Acknowledgements}

The authors would like to thank Dario Colombo for his help with the SCIMES Python package.
The Starlink software \citep{starlink} is currently sup-
ported by the East Asian Observatory. This research made use of
SCIMES, a Python package to find relevant structures into den-
drograms of molecular gas emission using the spectral clustering
approach \citep{Colombo2015}. SCIMES is an astropy affiliated-
package \citep{astropy:2013, astropy:2018}.

\section*{Data availability}

The data used for this paper are available from the archives of the
CHIMPS \citep{Rigby2019}. The SCIMES catalogue used for the analysis in this 
article is available to download from the CANFAR archive\footnote{\url{https://www.canfar.net/storage/list/AstroDataCitationDOI/CISTI.CANFAR/23.0003/data}}.



\bibliographystyle{mnras}
\bibliography{example} 




\appendix

\section{Segmentation of crowded and sparse fields}\label{exa}

\begin{figure*}
	\centering 
	\includegraphics[width=1.1\textwidth]{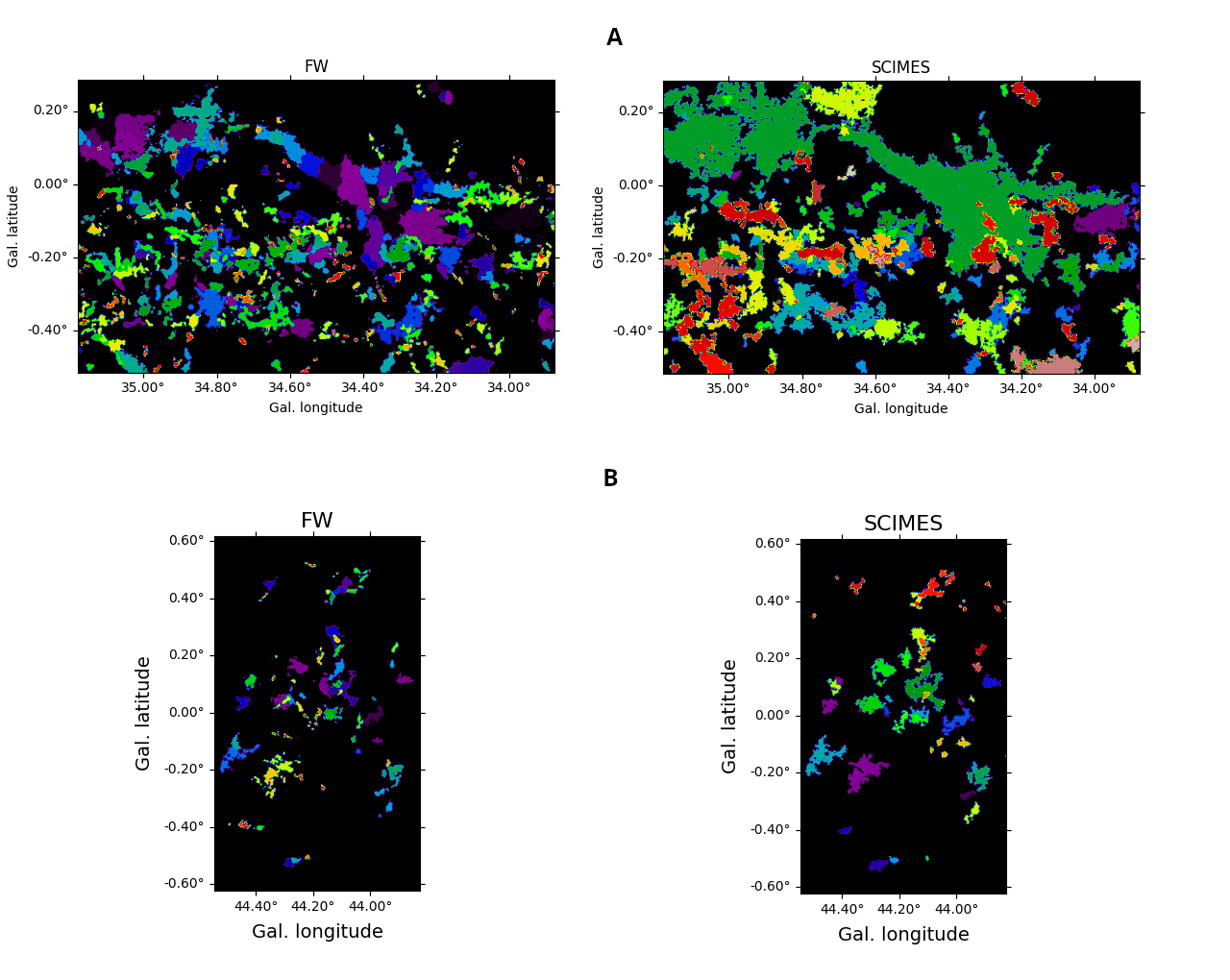} 
	\caption{Projected cloud assignments in the SCIMES and FW segmentations of region 3 (top panels) and region 9 (bottom panels) in CHIMPS. The clouds are colour-coded according to their assignment numbers in FW and SCIMES. In the case of overlapping clouds, the line-of-sight projection places the cloud with the highest assignment number on top. These projected maps provide an example of the performance of the SCIMES and FW algorithms in crowded and sparse fields (see Fig. \ref{rcuts}). } 
	\label{reg7} 
\end{figure*}

Fig. \ref{reg7} shows an example of FW and SCIMES segmentations of CHIMPS regions with different characteristics: one crowded (region 3) and one sparse field (region 9, see Fig. \ref{rcuts}).
In both situations, the average size of the sources identified by SCIMES is larger than those extracted by FW.

\section{Examples of SCIMES sources with no FW counterparts.}\label{exa2}

\begin{figure*}
	\centering 
	\includegraphics[width=1.1\textwidth]{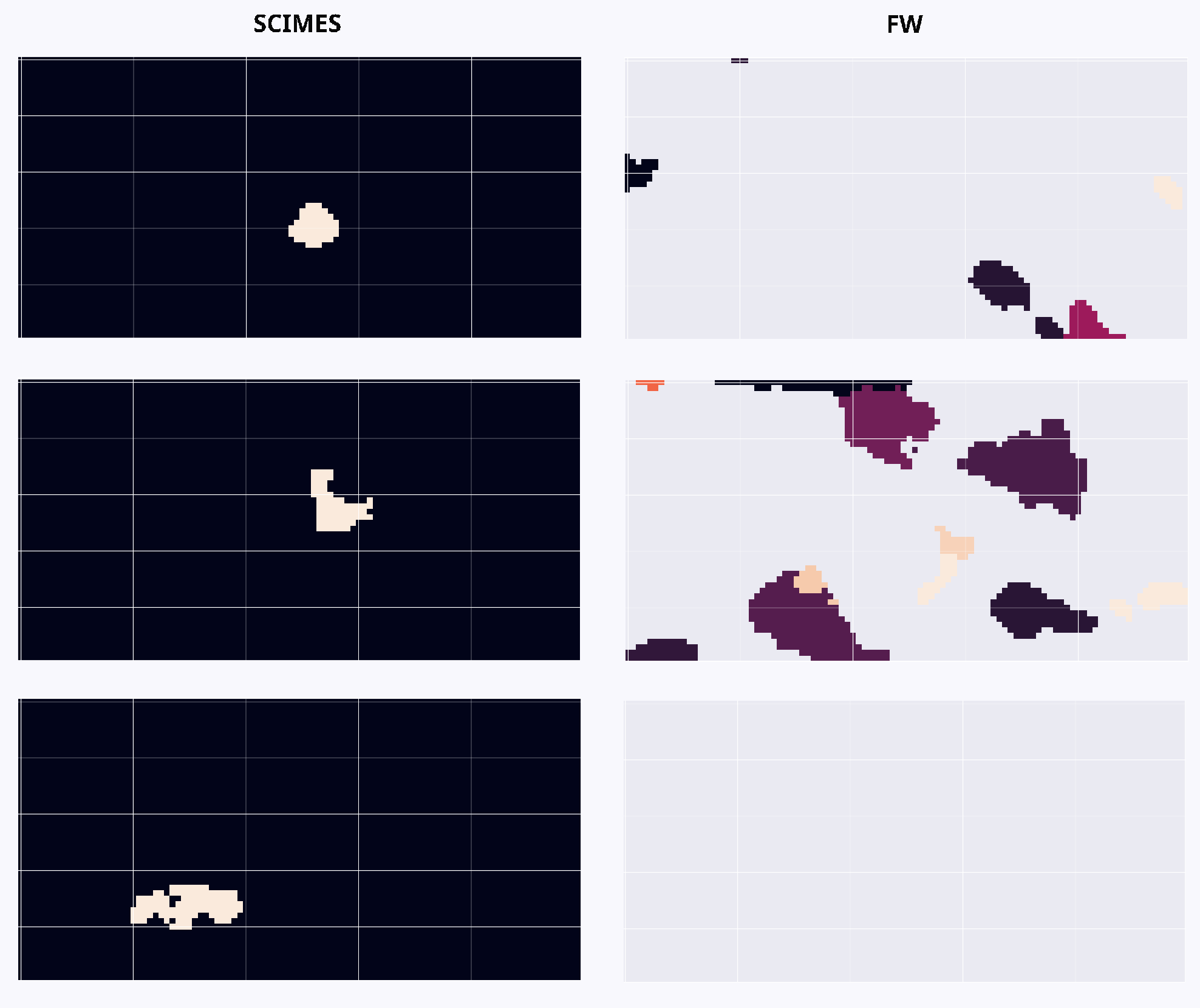} 
	\caption{Left column: projections along the spectral axis of 3 SCIMES sources that do not contained any emission peak found in the FW segmentation. Right column: the corresponding areas in the FW extraction. The projected sourcces are indicated by solid colors. } 
	\label{voxy} 
\end{figure*}

Fig. \ref{voxy} shows three examples of SCIMES sources that do not have a FW counterpart. As discussed in sub-section \ref{cloud_size}, these sources are more likely to be found in sparse fields and are smaller in volume (number of voxels) than the average SCIMES source. The current SCIMES extraction includes 540 such sources that contribute lower-than-average dispersion velocities to the overall distribution in the sample of SCIMES clouds, reducing the mean value.

%


\bsp	
\label{lastpage}
\end{document}